\documentclass[reprint,superscriptaddress,showpacs,floatfix
 amsmath,amssymb,
 aps,
prb,
]{revtex4-2}
\UseRawInputEncoding
\usepackage{graphicx}\usepackage{dcolumn}\usepackage{physics}
\usepackage{color}
\usepackage{mathtools}
\usepackage{bm}        \usepackage{amssymb}   \usepackage{amsmath}
\usepackage{wasysym}
\usepackage{comment}
\usepackage[colorlinks=true,citecolor=blue,linkcolor=magenta]{hyperref}
\clearpage{}\usepackage[normalem]{ulem}

\newcommand{\fig}[1]{Fig.~\ref{#1}}
\newcommand{\Fig}[1]{Fig.~\ref{#1}}

\newcommand{\eq}[1]{Eq.~\ref{#1}}

\makeatletter
\newsavebox{\@brx}
\newcommand{\llangle}[1][]{\savebox{\@brx}{\(\m@th{#1\langle}\)}\mathopen{\copy\@brx\kern-0.5\wd\@brx\usebox{\@brx}}}
\newcommand{\rrangle}[1][]{\savebox{\@brx}{\(\m@th{#1\rangle}\)}\mathclose{\copy\@brx\kern-0.5\wd\@brx\usebox{\@brx}}}
\makeatother

\newcommand{\br}{{\mathbf r}}

\newcommand{\bk}{{\mathbf k}}
\newcommand{\dg}{{\dagger}}
\newcommand{\pdg}{{\vphantom\dagger}}
\clearpage{}
\def \SI {\textcolor{blue}{\textit{SI Appendix}}}
\newcommand{\txt}[1]{\textcolor{blue}{SI~{#1}}}
\newcommand{\SIFig}[1]{\textcolor{blue}{Fig.SI~{#1}}}
\allowdisplaybreaks
\begin{document}

\preprint{APS/123-QED}

\title{Nematic phases and elastoresistivity
from a multiorbital non-Fermi liquid}

\author{Andrew Hardy}
\affiliation{Department of Physics, University of Toronto, 60 St. George Street, Toronto, ON, M5S 1A7 Canada}
\author{Arijit Haldar}\affiliation{Department of Physics, University of Toronto, 60 St. George Street, Toronto, ON, M5S 1A7 Canada}
\affiliation{S.N. Bose National Centre for Basic Science, JD-Block Sector III, Kolkata, India}
\author{Arun Paramekanti}
\email{arun.paramekanti@utoronto.ca}
\affiliation{Department of Physics, University of Toronto, 60 St. George Street, Toronto, ON, M5S 1A7 Canada}

\begin{abstract}
We propose and study a two-orbital lattice extension of the Sachdev-Ye-Kitaev model in the large-$N$ limit. 
The phase diagram of this model features a high temperature isotropic non-Fermi liquid
which undergoes a first-order thermal transition into a nematic insulator or a continuous thermal transition into 
a nematic metal phase, separated by a tunable tricritical point.
These phases arise from spontaneous partial orbital polarization 
of the multiorbital non-Fermi liquid. 
We explore the spectral and transport properties of this model,
including the d.c. elastoresistivity, which exhibits a peak 
near the nematic transition, 
as well as the nonzero frequency elastoconductivity. Our work 
offers a useful perspective on nematic phases and transport in correlated multiorbital systems.
\end{abstract}

\maketitle

The interplay of non-Fermi liquid physics (nFL)
with broken symmetry states of matter is a rich field of research 
in correlated electron systems. One approach to this
physics is to study metallic quantum critical points (QCPs) where 
fluctuations associated with the onset of symmetry breaking can
destroy quasiparticles on the Fermi surface \cite{millisEffectNonzeroTemperature1993,chubukovInstabilityQuantumCriticalPoint2004,Metlitski2010a,leeRecentDevelopmentsNonFermi2018,Raghu_PRL2019,bergMonteCarloStudies2019,VarmaRMP2020}.
An equally important exploration is to ask how nFLs, which may arise more generically in
correlated narrow-band materials including flat-band systems,
become unstable to diverse broken symmetry phases as we lower temperature \cite{Metlitski2015, Mandal2015, Mandal2016, ZhenInstability2017}.
Understanding these issues is of 
enormous interest for ongoing experiments on a wide range of 
quantum materials.

The electron nematic, a quantum liquid crystal, is a ubiquitous
broken symmetry phase associated with the loss of 
lattice rotational symmetry. Nematicity and nFL physics
have been explored in a host of correlated quantum 
materials including Moir\'e crystals such as twisted bilayer 
graphene with flat bands \cite{cao_nematicity_2021, cao_strange_2020,rubio-verdu_moire_2022},
iron pnictide and chalcogenide systems \cite{avci_magnetically_2014,fernandesWhatDrivesNematic2014,khasanov_magnetic_2018,Licciardello2019a, culo_putative_2021, Hirschfeld2021}, doped cuprates 
\cite{nie_vestigial_2017,mukhopadhyay_evidence_2019, gupta_vanishing_2021}, 
the bilayer strontium ruthenates \cite{Mackenzie_Science2007,kee2005,Raghu_PRB2009,Hayden_NatMat2015}, and
quantum Hall fluids
\cite{Eisenstein_Nematic2002,Fradkin2010a,Xia2011a, Feldman2016,Hayes2021a}.
Quantum criticality of uniform nematic order is also of great
interest since it impacts electrons on the entire Fermi 
surface.
Quantum Monte Carlo (QMC) simulations of sign-problem
free models show signatures of nFL properties
and an emergent superconducting
dome near such nematic QCPs \cite{schattner_ising_2016, bergMonteCarloStudies2019}.
On the experimental front, a particularly useful
tool to detect nematic fluctuations and symmetry
breaking is elastoresistivity. This measures the impact
of uniaxial strain on the resistive anisotropy,
providing a transport probe of the nematic susceptibility
\cite{kuoMeasurementElastoresistivityCoefficients2013,kuo_strain_science2016,mirri_electrodynamic_2016,palmstromComparisonTemperatureDoping2022,Hicks_PRX2021}.
While there has been progress in exploring elastoresistivity 
in strongly correlated Hubbard-type models \cite{arciniaga_theory_2020},
it is important to study extensions to
multi-orbital and multi-band
nFLs which are of relevance to diverse materials
including the cuprates,
FeSe, Sr$_3$Ru$_2$O$_7$, and Moir\'e crystals.

\begin{figure}
    \centering
\includegraphics[width=0.3\textwidth]{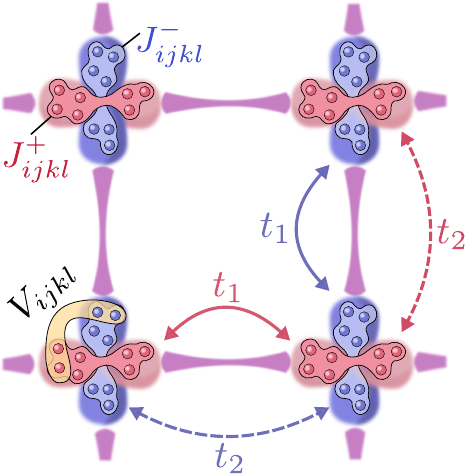}
    \caption{The two-orbital SYK model:
    Two independent SYK `dots' representing 
    orbitals
    (denoted as red ($+$) and blue-filled circles ($-$)) are placed 
    at each site $\br$ 
    of a square lattice. The orbitals at any given site have an SYK-type self-interaction
     (denoted $J^+_{ijkl}(\br)$ and $J^-_{ijkl}(\br)$) and an inter-orbital
    SYK-type interaction ($V_{ijkl}(\br)$). 
    Orbitals of the same type on neighboring lattice sites are connected via
    hopping terms. Each orbital has an `easy' axis 
    (hopping $t_1=t+\delta t$) and a `hard' axis (hopping $t_2=t-\delta t$).
    The easy axis for the `+' (red) orbitals points along the 
    horizontal direction, while the easy axis for the `$-$' (blue) orbitals 
    points along the vertical direction. The model is symmetric under a $C_4$ rotation, 
    provided we also exchange the orbital flavors.
    This symmetry is spontaneously broken in the nematic phase.
    }
    \label{fig:schematic}
\end{figure}
 Recently the theoretical study of nFLs has also seen significant progress. Starting with the formulation of the Sachdev-Ye-Kitaev (SYK) model 
 \cite{sachdev1993gapless, Sachdev2010, KitaevKITP, Maldacena2016} as a solvable example of nFL on a quantum dot, the field has grown to include several illuminating generalizations \cite{banerjeeSolvableModelDynamical2017c,Haldar2018c, esterlisCooperPairingIncoherent2019,wangSolvableStrongCouplingQuantumDot2020, lantagne2021,Chowdhury2021b}. 
 In this context, lattice extensions of the SYK model 
 are particularly interesting since they
 provide a controlled route to accessing 
 several phenomena, including lattice 
 nFLs \cite{gu2017local, Haldar2018b}, FL to nFL crossovers \cite{Song2017a, zhang2017,Chowdhury2018}, metal-insulator transitions \cite{jian2017model}, heavy fermion physics \cite{aldapeSolvableTheoryStrange2022}, and critical Fermi surfaces \cite{Chowdhury2018, esterlis2021}.
However, the question of how nematicity impacts a high-temperature nFL phase of an SYK lattice remains unexplored. 
In particular, is it possible to formulate a theoretically-solvable microscopic lattice model with 
strong interactions that spontaneously manifests nematic phases, and to study its transport properties?
 
Here, we address this question and other issues highlighted
above by constructing a two-orbital extension of a lattice
SYK model schematically depicted in \fig{fig:schematic}. 
The two orbitals may be viewed as representing, 
for instance, $d_{xz}$ and $d_{yz}$ orbitals, each with a 
preferred hopping direction, which play a role 
in many quantum materials. The underlying symmetry of this
system is a 
$C_4$ lattice rotation followed by the 
exchange of the two orbitals. Uniform orbital polarization breaks this
symmetry down to $C_2$, resulting in an Ising ferronematic, while
staggered orbital polarization results in an Ising antiferronematic.
Focusing on uniform orders, we study the complete phase diagram, 
thermodynamics, spectral functions, and transport, for
this model in the large-$N$ limit. We show that this model exhibits a
non-Fermi liquid phase at high temperature, which gives
way to a nematic insulator or a nematic metal upon cooling.
Depending on parameters, this thermal
transition is first-order or continuous, with a tunable
tricritical point.
We examine the transport properties of this model, including
the impact of uniaxial strain
which breaks orbital degeneracy. We find that uniaxial $B_{1g}$
strain leads to a peak in the d.c. elastoresistivity 
anisotropy in the vicinity of the isotropic-to-nematic transition, 
and we also present results on the frequency dependent elastoconductivity. 

Finally,
we extend our work to present preliminary results on checkerboard type 
antiferronematic order.

\section*{Two-orbital SYK lattice model}

The SYK model represents a single-site `dot' with $N$ 
fermionic modes having random all-to-all interactions which 
is exactly solvable when $N \!\to\! \infty$.
We generalize the SYK model to a square lattice
with each site having two `SYK dots', representing 
two orbitals, and each orbital accommodating 
$N$ fermionic modes. Modes in the two orbitals hop
anisotropically on the lattice, with a preferred direction 
as shown in 
\fig{fig:schematic}: modes in orbital $s\!=\! ``+"$ (red) 
hop along the
$\hat{x}$ and $\hat{y}$ directions with respective
amplitudes $t_1=t+\delta t$ and $t_2=t-\delta t$, with
$\delta t \!>\! 0$, and 
vice versa for modes in
orbital $s\!=\! ``-"$ (blue). The kinetic energy
is
\begin{equation}\label{eq:Hkin}
    H_{\rm kin} =  \sum_{\bk,s,i} \varepsilon^\pdg_s(\bk) c^{\dg}_{\bk,s,i} c^{\pdg}_{\bk,s,i}
\end{equation}
where $\bk$ is the momentum, $s\!=\! \pm$ is the orbital,
$i\!=\!1 \ldots N$ denote modes in each orbital, and the dispersion
$\varepsilon^\pdg_\pm (\bk) \! =\! -2 t (\cos k_x \!+\! \cos k_y) \!\mp\! 2 \delta t (\cos k_x \!-\! \cos k_y)$. The symbols $c^\dagger$, $c$ represent creation and annihilation operators for the fermions.

The interactions take on the SYK form, with two-body
intra-orbital and inter-orbital pair-hopping terms:
\begin{eqnarray}
\!\!\!\! H^{\rm intra}_{\text{SYK}}  &=& \!\!\!\!\!\! \sum_{\br,s,(ijkl)}
  \!\!\! J^{(s)}_{ijkl}(\br) 
  c_{\br,s,i}^{\dagger} c_{\br,s,j}^{\dagger} c^\pdg_{\br,s,k} 
  c^\pdg_{\br,s,l} \\
 \!\!\!\! H^{\rm inter}_{\text{SYK}} &=& \!\!\!\!\!\! 
  \sum_{\br,(ijkl)} 
  \!\!\! V_{ijkl}^\pdg(\br) 
  c_{\br,+, i}^{ \dagger}  c^{ \dagger}_{\br,+,j} 
  c^\pdg_{\br,-,k}  c^\pdg_{\br,-,l} + {\mathrm {H.c.}}
\end{eqnarray}
where ``H.c." denotes Hermitian conjugate and $\br$ denotes the position of a lattice site.
The couplings $J^{(s)}_{ijkl}(\br)$, $V_{ijkl}(\br)$ are uncorrelated random complex numbers having Gaussian distributions with \emph{zero}-mean,
and satisfy 
$\llangle {J^{(s)}_{ijkl}(\br)}^*J^{(s')}_{ijkl}(\br')\rrangle = \delta_{ss'}\delta_{\br\br'}J^2/(2N)^{3}$ and 
$\llangle {V_{ijkl}(\br)}^*V_{ijkl}(\br')\rrangle = \delta_{\br\br'}V^2/(2N)^{3}$, respectively. Here $\llangle\cdots\rrangle$ denotes the average over all disorder realizations and $J^2/(2N)^3$ and $V^2/(2N)^3$ sets the variance.
The couplings are properly antisymmetrized to obey
$J^{(s)}_{ijkl}=-J^{(s)}_{jikl}=-J^{(s)}_{ijlk}$, and a similar condition holds for $V_{ijkl}$ as well.

To solve for the phase diagram of this model, we go to the imaginary-time path
integral formulation and
average over disorder realizations for $J^{(s)}_{ijkl}$ and $V_{ijkl}$ via the replica trick. 
Doing so results in a disorder-averaged action parameterized by $J^2$ and $V^2$.
Next, we integrate out the fermion fields and rewrite the action using a replica-diagonal ansatz for the site-local imaginary-time Green's function $G_s(\tau) \equiv-(1/N)\sum_i\left\langle T_\tau c_{s,i}(\tau) c_{s,i}^{\dagger}(0)\right\rangle$ for each orbital $s$ and their corresponding self-energies $\Sigma_s(\tau)$ (see \SI,~\txt{1}). Here, $\tau$ represents the imaginary-time coordinate and $T_\tau$ is the time-ordering operator.

In the large-$N$ limit, the free energy functional $\Omega$ for our model can be obtained from the resulting action given by
\begin{equation}
\label{eq:freeenergy}
\begin{gathered}
\Omega =\sum_{s=\pm}\left[- \frac{1}{\beta} \sum_{i \omega_{n}} 
\int \mathrm{d} \varepsilon g_s(\varepsilon) \ln \left[\mathfrak{i} \omega_{n}  +\mu-\varepsilon-\Sigma_s\left(\mathfrak{i} \omega_{n}\right)\right]\right. \\
\left.+\int_{0}^{\beta} \mathrm{d} \tau \Sigma_s(\tau) G_s(\beta-\tau) - \frac{J^{2}}{4} \int_{0}^{\beta} \mathrm{d} \tau 
G_s^{2}(\beta-\tau) G_s^{2}(\tau)\right]  \\
-  \frac{V^2}{2} \int_{0}^{\beta} d \tau G_{+}^{2}(\beta-\tau) G_{-}^{2}(\tau)
\end{gathered}
\end{equation}
where $g_s(\varepsilon) = \int \frac{d^2 {\bk}}{(2\pi)^2} \delta(\varepsilon-\varepsilon_s(\bk))$ 
is the lattice density of states for orbital-$s$, $\mu$ is the chemical potential,  $\omega_n\!=\! (2 n \!+\! 1)\pi/\beta$ 
represents the fermionic Matsubara frequencies, and $\beta=T^{-1}$ with $T$ denoting the temperature.

We note here since the dispersions for the $+$,$-$ orbitals obey $\varepsilon_+(k_x,k_y)$=$\varepsilon_-(k_y,k_x)$ (see \eq{eq:Hkin}), both orbitals are described by the same density of states, so that $g_+(\varepsilon) = g_-(\varepsilon) = g(\varepsilon)$ 
in \eq{eq:freeenergy}. However, despite having the same $g(\varepsilon)$, the inter-orbital SYK-interaction $V$ can still drive a spontaneous symmetry breaking between the orbitals, as we demonstrate in the next section.
The imaginary-time Green's function $G_s(\tau)$ in \eq{eq:freeenergy} satisfy the boundary condition $G_s(-\tau)\!=\! -G_s(\beta-\tau)$, and so does the imaginary-time self energy $\Sigma_s(\tau)$. 
Setting $\delta\Omega/\delta G_s(\tau)\!=\! 0$ and
$\delta\Omega/\delta\Sigma_s(i\omega_n)\!=\!0$
leads to 
\begin{eqnarray}
\label{eq:selfcon1}
\Sigma_s(\tau)&=& -J^2 G_s^2(\tau) G^\pdg_{s}(-\tau)-V^2 G^2_{-s}(\tau) G^\pdg_{s}(-\tau)\\
\label{eq:selfcon2}
G_s(i\omega_n)&=& \int \mathrm{d} \varepsilon g(\varepsilon) \left[i\omega_n+\mu-\varepsilon-\Sigma_s(i\omega_n)\right]^{-1}
\end{eqnarray}
which we solve self-consistently
(see Ref.~\cite{Haldar2018b} and \SI,~\txt{3}). 
The solution is 
used to compute the equilibrium free energy and thermodynamic properties using \eq{eq:freeenergy}.
While these equations have been obtained starting from the SYK model in the large-$N$ limit, 
we may also view them as
a type of self-consistent dynamical mean field theory of a two-orbital model (although we
caution that the lattice SYK equations, and results, differ from iterated perturbation theory 
for solving the Hubbard model within dynamical mean field theory (DMFT) \cite{Georges1996,Held2007}).

\section*{Phase Diagram}
\begin{figure*}
\centering
\includegraphics[width=\textwidth]{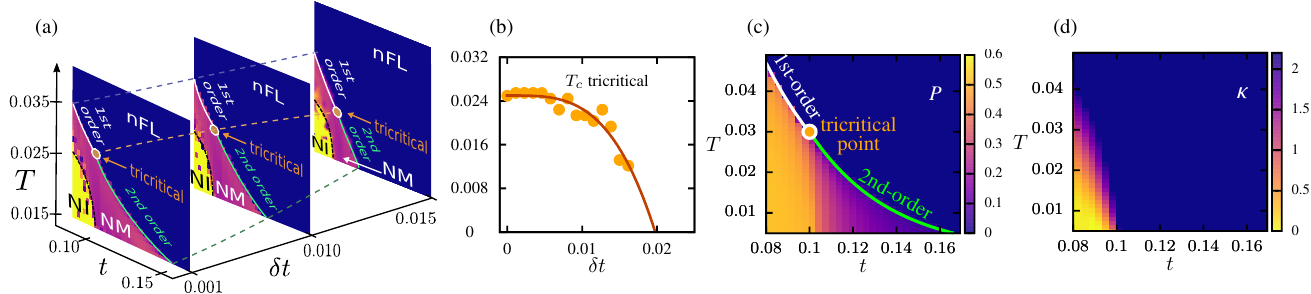}
\caption{(a) Phase diagram in terms of temperature ($T$) and hoppings ($t$ and $\delta t$), showing an isotropic non-Fermi liquid (nFL),
a nematic metal (NM), and a nematic insulator (NI). We have set 
$J\!=\!V\!=\!1$ and the density to half-filling.
The isotropic and nematic
phases are separated by first-order or continuous thermal transitions which meet at a tricritical point (filled circle). The NM and NI
regimes are separated by a crossover at nonzero $T$. 
(b) Temperature at the tricritical point versus
$\delta t$ showing that it could be potentially further 
tuned (see solid line, guide to eye) to reach a quantum tricritical point in a more generalized model.  (c) Polarization ($P$), or orbital
density imbalance for a cross section taken along the $t$-$T$ plane in (a) with $\delta t/t=0.05$. 
(d) Compressibility ($\kappa$), distinguishing metallic (nFL and NM) from insulating 
phases (NI), for a $t-T$ plane cross section with $\delta t/t=0.05$.}

\label{fig:phasediag}
\end{figure*}

We begin by discussing the uniform nematic order
which appears as a symmetry-breaking solution 
to the above equations
with $\Sigma_{+}\neq\Sigma_{-}$. This phase, driven by the 
inter-orbital interaction $V$,
may be viewed as a lattice generalization of the
flavor-imbalanced phase of a two-flavor SYK model \cite{Haldar2018c, HaldarPRR2020}. The inter-orbital $V$ interaction hops a 
pair of electrons from modes of a single orbital $s$ 
to modes of the other orbital $\bar{s}$, and is 
thus distinct from the
original single SYK dot interaction which randomly hops 
electrons between
any two pairs of fermionic modes. While the latter does not
have a symmetry broken phase, the $V$ interaction can indeed induce
symmetry breaking.
As shown in \fig{fig:phasediag}(a), for $V\!=\!J\!=\!1$,
the nematic phase is separated from an isotropic 
nFL by a first-order nematic transition at 
small hopping amplitude $t$. Increasing $t$, we encounter a tricritical
point beyond which the thermal-nematic transition becomes 
second order. 
The tricritical point can be tuned by the
hopping anisotropy $\delta t$ as shown in \fig{fig:phasediag}(b).

We note that the nematic phases appear in the regime $V\! \gg\! t$ which is the regime certainly
relevant to flat-band systems. This is also reasonable for typical correlated oxides or chalcogenides
where the hopping $t=100$-$300$ meV while the scale of electron-electron interactions is $\sim 1$-$3$eV.
We have also explored how the phase diagram changes as we tune $V/J$; increasing $V/J >\! 1$ does not qualitatively
modify our results but leads to quantitative shifts in the phase boundaries, where the nematic transition 
occurs at higher temperatures as $V/J$ increases. This is explored in more detail in \SI{},~\SIFig{2},~\SIFig{3}, and \txt{3}.

We characterize the nematic 
order by the orbital polarization $P\!=\! \langle n_+ 
\rangle \!-\! \langle n_{-} \rangle$, where the orbital
densities are computed as 
$\langle n_s \rangle \!=\! G_s(\tau \!=\! 0^-)$. 
As seen from \fig{fig:phasediag}(c), $P$ increases
sharply below the first-order transition, while it increases
gradually below the continuous thermal transition. 
Everywhere in the nematic phase, the polarization remains 
below its maximal value $P_{\rm max}\!=\! 1$. These transitions
also exhibit corresponding signatures in the entropy
$S \!=\! - \partial\Omega/\partial T$ and the specific heat 
$C_v \!=\! T \partial S / \partial T$ (\SI,~\SIFig{1}).
The existence of a tricritical point in the phase diagram 
can be qualitatively understood 
using a Landau Ginzburg (LG) theory type approach as discussed in 
\SI ~\txt{7}; also see previous work on two coupled SYK dots  \cite{Haldar2018c}.
In the Outlook, we discuss closely competing staggered
nematic phases which result from a more challenging numerical
solution of the lattice SYK equations on an enlarged unit cell,
in a spirit similar to cellular DMFT 
\cite{Kotliar2001,Capone2006}.

We next discuss the behavior of the electronic compressibility
$\kappa \!=\! \langle n \rangle^{-2} \pdv[2]{\Omega}{\mu} = \langle n \rangle^{-2} \pdv{\langle n \rangle }{\mu}$,
where, $\langle n\rangle=- \partial\Omega/\partial\mu$ is the total number density of fermions  set to half-filling by particle-hole symmetry $(\langle n \rangle = 0.5)$.
As seen in \Fig{fig:phasediag}(d), the compressibility vanishes
as $T \to 0$ at small $t$, but it remains nonzero at larger $t$.
This allows us to distinguish insulating from metallic phases,
which we also confirm below using spectral functions and
transport. We observe that the nematic phase exhibits both
metallic (NM) and insulating (NI) regimes; these appear
to be separated by a continuous transition as $T\!\to\! 0$,
and by the indicated crossover lines in 
\fig{fig:phasediag}(a) at nonzero $T$. 

\section*{Spectral functions}

To explore the spectral properties across the phase diagram,  we have analytically continued the self-consistent equations \ref{eq:selfcon1} and \ref{eq:selfcon2} from the Matsubara frequencies ($i\omega_n$)  to the real frequency line ($\omega\in\mathbb{R}$) following the method presented in Ref. \cite{banerjeeSolvableModelDynamical2017c,Haldar2018b} (also see \SI, \txt{2} for a summary). Solving these real-tim equations on the real frequency axis allows us to extract the onsite retarded Green's functions $G^{R}_s(\omega)$ and the retarded self-energies $\Sigma^{R}_s(\omega)$ without resorting to any numerical algorithms for analytic continuation, such as Pad\'{e} approximation or maximum entropy. We then utilize $G_s^R(\omega)$ to compute the local (onsite) spectral function $A_s(\omega)$ for each orbital  using $A_s(\omega) \!=\! -{\rm Im} G^{R}_s(\omega)/\pi$.

\begin{figure}[t]
\includegraphics[width=0.5\textwidth]{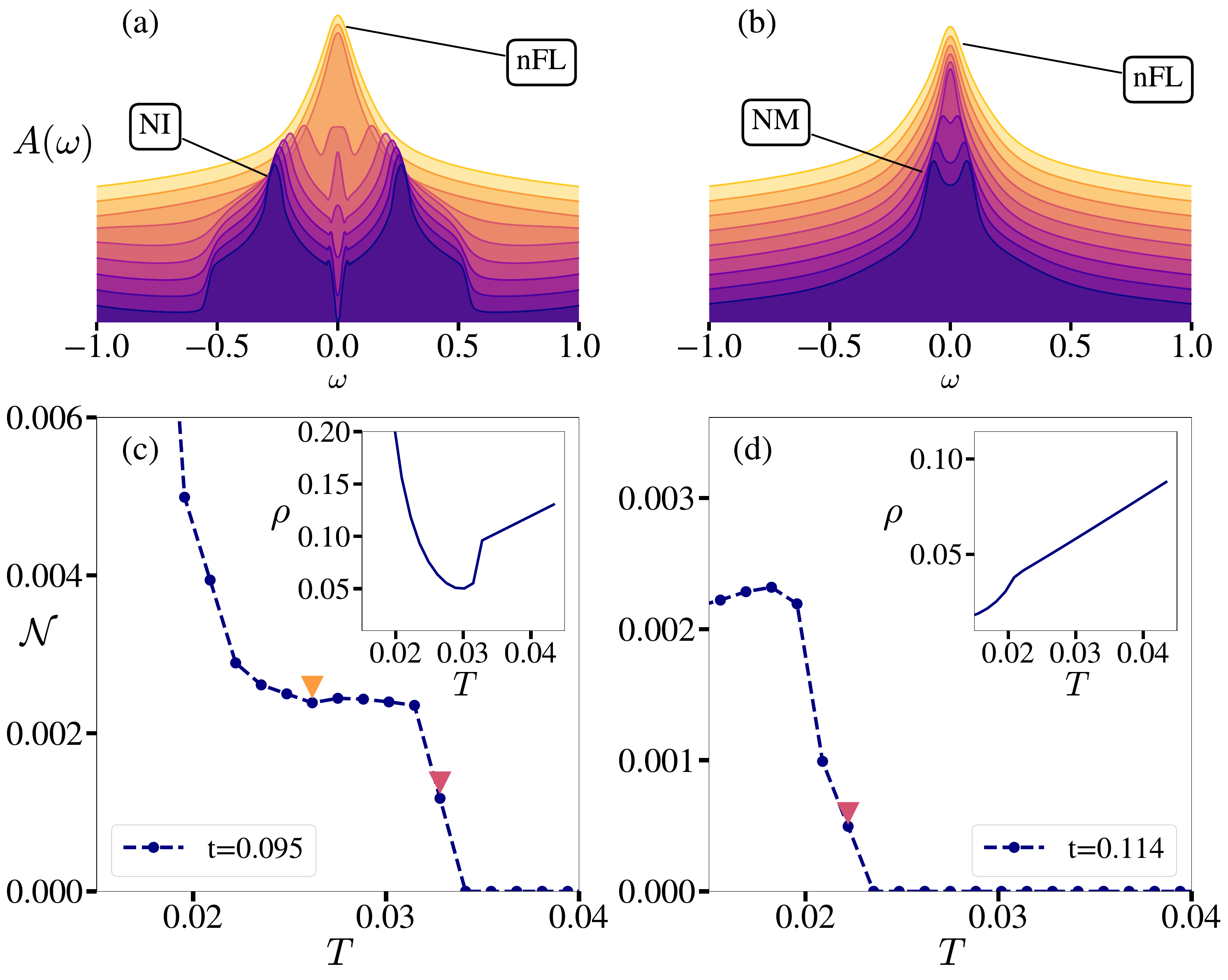}
    \caption{(a) Evolution of the local spectral function $A(\omega)$ with temperature $T$ as a 
    function of frequency $\omega$ upon cooling from the isotropic nFL into the NI.  We find an intermediate temperature regime
    where nematicity coexists with a peak at $\omega\!=\!0$ is 
    indicative of the NM. Upon cooling further,
    this peak converts into a gap, leading to the NI. (b) Same as (a) but for a regime where
the $T \to 0$ phase is a NM, indicated by a finite-spectral weight at $\omega=0$.
    (c) Resistive nematicity ${\cal N}(T)$ which
    is zero in the isotropic phase, increases (purple arrow) and
    saturates in the NM, and further rapidly increases (yellow arrow) at low $T$ in the NI regime. Inset shows the average
    resistivity $\rho(T)=(\rho_{xx}+\rho_{yy})/2$.
    (d) Same as (c), but retains the NM state upon cooling. Inset indicates corresponding average resistivity 
    $\rho(T)$. $\delta t/t = 0.05$ for all subfigures.}
    \label{fig:dynamics}
\end{figure}

\Fig{fig:dynamics}~(a,b)
shows the total 
spectral function $A(\omega)\!=\! \sum_s A_s(\omega)$ in the regimes
where the $T=0$ phase is (a) a nematic insulator and (b) a nematic
metal respectively. At high $T$, both regimes feature an orbital
symmetric nFL regime with a peak at $\omega\!=\! 0$. Since
orbital symmetry breaks below the nematic transition, the 
spectral functions for the two orbitals split and move away from
$\omega\!=\!0$, leading to twin peaks in 
$A(\omega)$, one from each orbital. 
Since $\langle n_s \rangle \!=\! \int_{-\infty}^{\infty}\!d\omega\ 
n_F(\omega) A_s(\omega)$, this splitting correlates with the onset
of nonzero polarization $P$ discussed earlier. Here, $n_F(\omega)$
denotes the Fermi function. The symmetries of the local spectral function
in the isotropic phase, which include particle-hole symmetry, 
$A_s(\omega)=A_s(-\omega)$, and orbital-exchange symmetry,
$A_s(\omega)=A_{-s}(\omega)$, get reduced in the nematic phase
to a combined symmetry under $\omega\!\to-\omega$ followed by orbital 
exchange $s\to-s$, so that $A_s(\omega)\!=\! A_{-s}(-\omega)$.

From \fig{fig:dynamics} (a), we find a regime of temperatures below the nematic transition
where a spectral peak survives at $\omega\!=\!0$, suggestive of
a NM, before a (soft) gap opens up leading to a 
loss of low frequency spectral weight, and eventually a
hard insulating gap at $T=0$.
This is consistent with a vanishing compressibility
$\kappa(T \!\to\! 0)$ for small values of hopping $t$ in \fig{fig:phasediag}(d). \\
Furthermore, since $\Sigma^R(\omega)$ is $\bk$-independent, the momentum-resolved spectral function
$A_s(\bk,\omega)$ is obtained from the lattice Green's function
$G^R_s(\bk, \omega)\!=\! 1/(\omega+\mu-\varepsilon_s(\bk)\!-\!\Sigma^R_s(\omega))$ using $A_s(\bk,\omega)\!=\! -\Im G^R_s(\bk,\omega)/\pi$. Representative plots for $A_s(\bk,\omega)$ and related details are given in \SI,~\txt{4} and \SIFig{4}.

We next turn to the quasiparticle residue $Z$ and effective mass $m^*$, dropping the orbital label
since these observables turn out to be the same for both orbitals. 
We define $Z(T) = \left(1 - \pdv{\Sigma(\omega)}{\omega}\right)^{-1}\!|_{\omega \rightarrow 0}$, so that the 
quasiparticle residue in a Fermi liquid ground state corresponds to $Z(T \!\to \! 0)$. 
Since the self-energy in our model is momentum-independent, the temperature dependent effective mass enhancement
may be written as $m^*(T) / m  = Z^{-1}(T)$. As discussed in detail in \SI~\txt{4}, we find that upon cooling the isotropic nFL,
$m^*(T)/m$ grows and appears to diverge. However, this growth is cut-off below the nematic transition, leading to a
finite mass enhancement and a correspondingly reduced $Z < 1$.

\section*{Transport}

Given the spectral functions above,
the real part of the conductivity (per flavor, i.e. scaled by $1/N$) is computed as
(along both directions $\alpha=x,y$)
\begin{eqnarray}
\label{kubo}
{\mathrm{Re}}\sigma_{\alpha\alpha}(\omega, T) \!&=&\! \frac{1}{\omega} \! \sum_s \int_{\bk,\omega'} \!\!\! v_{s,\alpha}^{2}(\bk) A_s\left(\bk,\omega^{\prime}\right) A_s\left(\bk,\omega\!+\!\omega^{\prime}\right) \nonumber \\
&\times& \left[n_F\left(\omega^{\prime}\right)-n_F\left(\omega+\omega^{\prime}\right)\right],
\end{eqnarray}
where $\vec{v}_{s}(\bk)\!=\! \vec \nabla_{\bk} \varepsilon_s(\bk)$.

In the $N \!\rightarrow \!\infty$ limit, this disorder averaged 
Kubo formula result has been shown to be exact, with no vertex 
corrections \cite{patelTheoryPlanckianMetal2019, joseNonFermiLiquidBehavior2022}, and with a
separable product of disorder averaged 
Green functions $\overline{GG} = \bar{G} \bar{G}$ \cite{Chowdhury2018}. We extract the d.c. conductivity as
the slope of the $\omega \sigma_{\alpha\alpha}(\omega)$ curve
for $\omega\!\to\!0$,
and invert it to obtain the d.c. resistivity 
$\rho_{\alpha\alpha}$. \Fig{fig:dynamics}(c) and (d) show
the \emph{resistive nematicity} ${\cal N} = (\rho_{xx}-\rho_{yy})/(\rho_{xx}+\rho_{yy})$ as a function of
temperature, with the average resistivity plotted in the corresponding insets. 
The high temperature nFL regime displays a characteristic $\rho \propto T$ 
resistivity. This nFL phase is a well-known result common to high-temperature lattice-SYK 
models \cite{Chowdhury2021b}. As we enter the nematic phase at low
temperature, the decrease in the spectral weight at the Fermi level (see \Fig{fig:dynamics}(b)) 
cuts off the effect of strong scattering from the SYK interactions, and leads to a FL with $\rho \propto T^2$.
We find that ${\cal N}(T)$ vanishes
in the isotropic phase, while it displays
a plateau in the 
NM, before rapidly increasing to 
${\cal N } \!\sim\! {\cal O}(1)$ 
deep in the NI.

\section*{Elastotransport}

\begin{figure}
\centering
\includegraphics[width=0.5\textwidth]{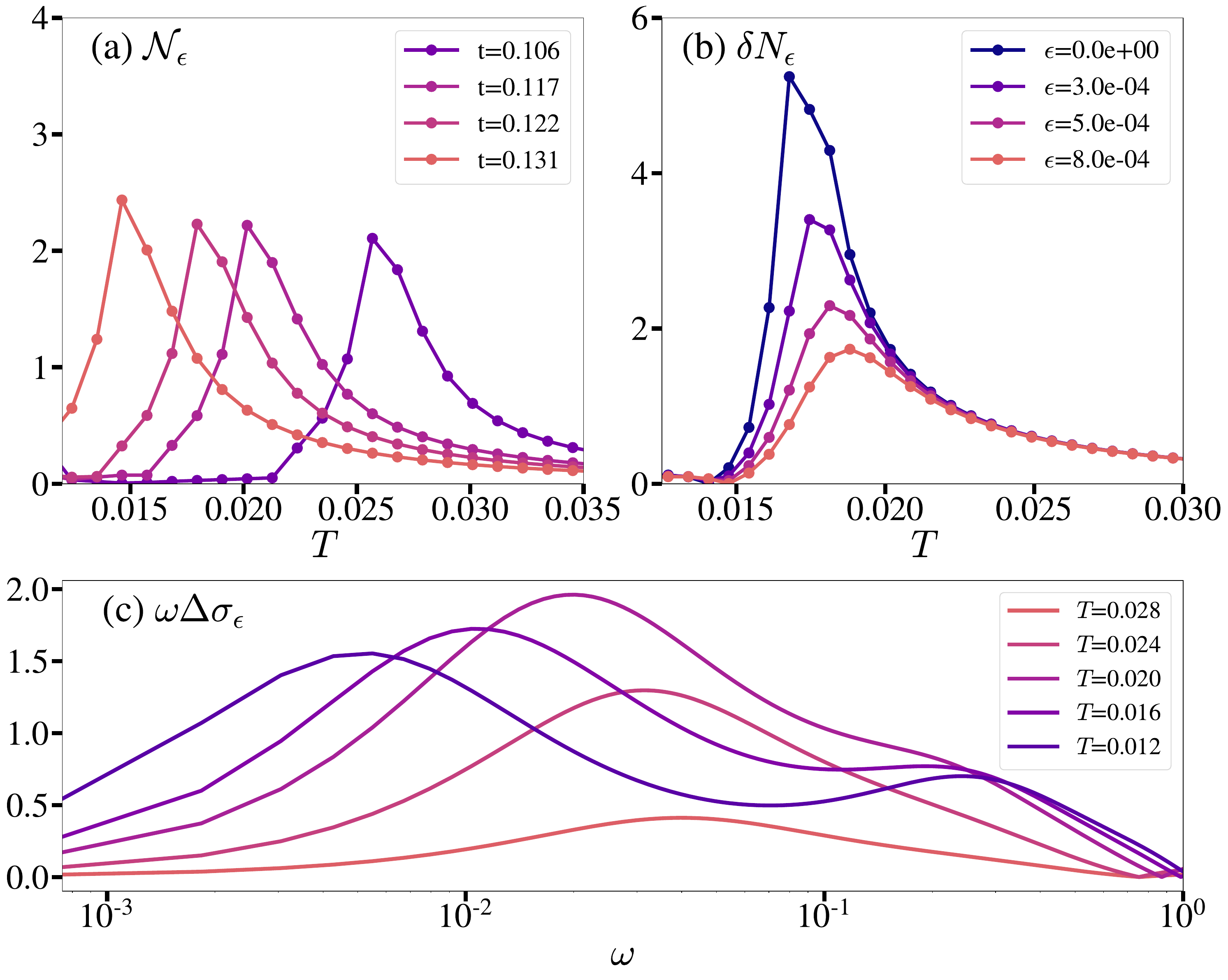}
\caption{ (A) ${\cal N}_\epsilon(T)=(\delta \rho_{xx}-\delta\rho_{yy})/\epsilon$ is the anisotropic 
differential elastoresistivity 
(see text for details) for $\epsilon = 10^{-3}$,
for various cuts through the nFL-NM transition; the peaks correspond 
to $T_c$. (b) Strain dependence of ${\delta \cal N}_\epsilon(T)$ for $t = 0.12$ showing
that the transition and hence ${\delta \cal N}_\epsilon(T)$
gets rounded with increasing strain. (c) 
$\omega \Delta \sigma_\epsilon(\omega)$ versus frequency for 
$t = 0.12$, 
where $\Delta \sigma_\epsilon(\omega)$ is the anisotropic differential elastoconductivity, 
at fixed small strain $\epsilon=10^{-3}$. Here, $\delta t/t = 0.05$ for all subfigures. 
}
\label{fig:elasto}
\end{figure}

It has 
been shown that transport in the presence of uniaxial strain 
can provide a sensitive probe of nematic fluctuations and
the onset of nematic order \cite{mirri_electrodynamic_2016,arciniaga_theory_2020}. In order to explore this, 
we assume that the uniaxial strain imposes a local orbital 
splitting which varies linearly with the strain; physically,
this will arise due to a modification in the local crystal 
field environment. Given its strain-induced origin, we use
$\epsilon$ to denote this splitting. For $B_{1g}$ strain, 
this leads to an additional term in the Hamiltonian
$\epsilon \sum_\br (n_{\br,+,i}-n_{\br,-,i})$ which explicitly breaks
the $C_4$ symmetry by favoring one of the two orbitals. To explore
the impact of strain on transport, we solve the 
self-consistent large-$N$ equations with a nonzero 
$\epsilon$ and compute changes in the resistivity $\delta \rho_{\alpha\alpha}$.
\Fig{fig:elasto}(a) shows the computed differential 
anisotropic elastoresistivity \cite{kuoMeasurementElastoresistivityCoefficients2013}
\begin{equation}
{\cal N}_\epsilon= \frac{1}{\epsilon}\left( \frac{\delta \rho_{xx}(\epsilon)}{ \rho_{xx}(0)} -\frac{\delta\rho_{yy}(\epsilon)}{ \rho_{yy}(0)}\right), 
\end{equation}
for different hoppings $t$, corresponding to different
cuts through the phase diagram which 
pass through the nematic metal. Here $\delta\rho_{\alpha\alpha} = (\rho_{\alpha\alpha}(\epsilon) - \rho_{\alpha\alpha}(0) ) $ represents
the change in resistivity from its unstrained value
due to a weak nonzero $\epsilon=10^{-3}$ . We
find that ${\cal N}_\epsilon(T)$ shows a significant
increase upon cooling towards the nematic transition, with a peak 
at the onset of nematic order. In this particle-hole
symmetric model, we expect ${\cal N}_\epsilon$ to be dominated
by changes in the orbital occupation rather than inducing
orbital-dependent scattering rates, so that ${\cal N}_\epsilon$ 
reflects changes in the orbital
polarization due to strain, and is thus tied to the nematic susceptibility.
\Fig{fig:elasto}(b) shows the strain dependence of the 
differential anisotropic elastoresistivity 
$\delta{\cal N}_\epsilon(T)$
for a fixed hopping $t$,
where we compute the resistivity $\delta\rho_{\alpha\alpha}$ 
at two nearby strain values $\epsilon$ (indicated
in the plot) and $\epsilon+d\epsilon$
with $d\epsilon=10^{-4}$. We
see that with increasing $\epsilon$, which imposes orbital
symmetry breaking, the nematic phase 
transition gets rounded out. These results on elastotransport
bear a striking resemblance to experimental observations on the
iron-based materials \cite{kuo_strain_science2016,palmstromComparisonTemperatureDoping2022}.

We have also studied the effects of $B_{2g}$ strain on our model by introducing a hybridization 
term $H_{\gamma} =  \gamma \sum_{s} c^\dagger_{s} c^{}_{-s} $ in the Hamiltonian (\eq{eq:Hkin}). 
We find that this off-diagonal term acts as a  
transverse field on
the Ising orbital order 
\cite{maharajTransverseFieldsTune2017}, which can tune the nematic transition and tricritical point to lower 
temperatures potentially leading
to quantum critical and tricritical points (see \SI,~\SIFig{6}). Additional details of this 
analysis and other results are discussed in \SI,~\txt{5}.

We finally turn to the frequency-dependent elastoconductivity
for weak nonzero $B_{1g}$ strain.
\Fig{fig:elasto}(c) shows plots of
$\omega \Delta\sigma_\epsilon(\omega)$
 as a function of temperature as we cool into the
 nematic insulator, where $\Delta\sigma_\epsilon(\omega)\!=\!
 {\mathrm{Re}}(\delta\sigma_{xx}-\delta\sigma_{yy})/\epsilon$ 
 is the differential anisotropic elastoconductivity 
 obtained from the change in conductivities due to $\epsilon=10^{-3}$.
We find that $\omega \Delta\sigma_\epsilon(\omega)$ exhibits a bump
near $\omega \!\sim\! 0.03$ (at higher $T$), which
shifts to lower frequency upon cooling, and is largest
near $T_c$. We expect the location of this 
peak to track the scattering rate, with the peak height
tracking the nematic susceptibility.
Indeed, in a simple Drude-like theory, with ${\mathrm{Re}}\sigma(\omega) 
 =(ne^2\tau/m^*)/(1+\omega^2\tau^2)$, 
 $\omega {\mathrm{Re}} \sigma(\omega)$ peaks at $\omega=1/\tau$, with the peak height $ne^2/2m$
 which is independent of the scattering rate $1/\tau$. 
 Here, $n$, $e$, $\tau$, and $m^*$ refer to the carrier density, charge, lifetime, and 
 effective mass for electrons.
It is thus plausible that the peak in $\omega \Delta \sigma_\epsilon(\omega)$ 
could also be a better measure of the nematic susceptibility, being independent of the 
scattering rate even in a more general setting where particle-hole
symmetry is lost. The second peak in  $\omega \Delta \sigma(\omega)$,
visible at higher frequency, reflects subtle features in the 
single-particle spectrum.

\section*{Outlook}

We have proposed a two-orbital model for a nFL which undergoes a $C_4 \!\to\! C_2$ rotational symmetry-breaking transition to a nematic phase, 
and studied its thermodynamic, spectral, and transport 
properties. We
have unveiled a rich phase diagram, where tuning the temperature
and orbital anisotropy in the nFL leads to nematic
critical or 
tricritical points. In addition, we have shown that strain
fields with different symmetries ($B_{1g}$ or $B_{2g}$) 
can be used to suppress the nematic ordering, potentially driving quantum critical or 
tricritical points in the nFL. Our results on the impact of strain on transport
in a nFL are of broad interest for a wide range of quantum materials
such as twisted bilayer graphene, iron-based superconductors, and 
underdoped cuprates.
Our d.c. elastotransport results show a strong peak in the
elastoresistivity as we approach the nematic 
transition that closely
resembles experimental data obtained for the normal state of the
iron chalcogenide superconductors \cite{kuo_strain_science2016,palmstromComparisonTemperatureDoping2022,Hicks_PRX2021}. Furthermore, our work goes beyond a Boltzmann equation treatment of quasiparticles with impurity scattering near nematic symmetry breaking 
transitions which is only applicable in the weakly correlated regime \cite{Andersen_defect2014,Benfatto2022}. Therefore, our work can serve as a 
useful point of comparison against experiments as well as numerical
techniques capable of accessing the strongly-correlated regime, such as QMC studies of metals undergoing nematic ordering
\cite{schattner_ising_2016, bergMonteCarloStudies2019} 

Our predictions for the frequency-dependent elastoconductivity could be tested in future experiments; since the peak in $\omega \Delta \sigma_\epsilon(\omega)$ occurs at the scattering rate, exploring this physics may call for new THz spectroscopic probes in strained quantum materials.

We have also explored competing orders in our two-orbital SYK model via a numerical study using an expanded 
unit-cell, in the spirit of cellular DMFT 
\cite{Kotliar2001,Capone2006}. 
Since this is a far more challenging numerical computation, we have thus far 
only explored a limited set of parameters. We find that
a staggered nematic metal state, with a `checkerboard' pattern of orbital polarizations, 
is nearly degenerate with the uniform ferronematic,
but with a free energy density which is very slightly lower $\Delta \Omega \propto t^4/V^3$.
Applying even a small $B_{1g}$ strain, with $\epsilon/t \!\sim\! 10^{-3}$,
already tilts the balance in favor of the uniform ferronematic. These results are presented and discussed in more detail in \SI,~\txt{6} and \SIFig{7}.
Thus, even if the uniform nematic is
a metastable state for the limited set of parameters we have explored, very small modifications to the
Hamiltonian may be sufficient to render it as the stable ground state. Moreover, going to larger clusters may also
impact this competition. Resolving this issue is a topic for future investigation.

Turning to other future
research directions, an important question is a careful theoretical understanding
of strain-induced nematic quantum critical or tricritical
points in the background of the nFL.
Another important direction is to study the fate of 
nematic phases 
and phase transitions when the SYK couplings in our multiorbital
model are made translationally invariant at the microscopic Hamiltonian 
level rather than in a disorder averaged sense. This will lead to momentum-dependent 
self energies, which could be of potential interest for exploring pseudogap physics in
multiorbital systems.

\acknowledgments{
We thank Anjishnu Bose,
Ian Fisher, and Andrey Chubukov for useful discussions.
We acknowledge funding from the Natural Sciences and Engineering Research Council (NSERC) of Canada. 
A. Hardy acknowledges funding from the Canada Graduate Scholarship (CGS-M NSERC).
Numerical computations were performed on the Niagara supercomputer at the SciNet HPC Consortium and Compute Canada.
}

\noindent {\bf Author Information}

\noindent The authors declare no competing financial interests. Correspondence should be addressed to A.P. (\href{mailto:arun.paramekanti@utoronto.ca}{arun.paramekanti@utoronto.ca}). \\[-0.5mm]

\noindent{\bf Data availability}

\noindent  All study data are included in this article and \SI{}. \\

\bibliography{mainbib}

\begin{thebibliography}{70}%
\makeatletter
\providecommand \@ifxundefined [1]{%
 \@ifx{#1\undefined}
}%
\providecommand \@ifnum [1]{%
 \ifnum #1\expandafter \@firstoftwo
 \else \expandafter \@secondoftwo
 \fi
}%
\providecommand \@ifx [1]{%
 \ifx #1\expandafter \@firstoftwo
 \else \expandafter \@secondoftwo
 \fi
}%
\providecommand \natexlab [1]{#1}%
\providecommand \enquote  [1]{``#1''}%
\providecommand \bibnamefont  [1]{#1}%
\providecommand \bibfnamefont [1]{#1}%
\providecommand \citenamefont [1]{#1}%
\providecommand \href@noop [0]{\@secondoftwo}%
\providecommand \href [0]{\begingroup \@sanitize@url \@href}%
\providecommand \@href[1]{\@@startlink{#1}\@@href}%
\providecommand \@@href[1]{\endgroup#1\@@endlink}%
\providecommand \@sanitize@url [0]{\catcode `\\12\catcode `\$12\catcode
  `\&12\catcode `\#12\catcode `\^12\catcode `\_12\catcode `\%12\relax}%
\providecommand \@@startlink[1]{}%
\providecommand \@@endlink[0]{}%
\providecommand \url  [0]{\begingroup\@sanitize@url \@url }%
\providecommand \@url [1]{\endgroup\@href {#1}{\urlprefix }}%
\providecommand \urlprefix  [0]{URL }%
\providecommand \Eprint [0]{\href }%
\providecommand \doibase [0]{https://doi.org/}%
\providecommand \selectlanguage [0]{\@gobble}%
\providecommand \bibinfo  [0]{\@secondoftwo}%
\providecommand \bibfield  [0]{\@secondoftwo}%
\providecommand \translation [1]{[#1]}%
\providecommand \BibitemOpen [0]{}%
\providecommand \bibitemStop [0]{}%
\providecommand \bibitemNoStop [0]{.\EOS\space}%
\providecommand \EOS [0]{\spacefactor3000\relax}%
\providecommand \BibitemShut  [1]{\csname bibitem#1\endcsname}%
\let\auto@bib@innerbib\@empty
\bibitem [{\citenamefont {Millis}(1993)}]{millisEffectNonzeroTemperature1993}%
  \BibitemOpen
  \bibfield  {author} {\bibinfo {author} {\bibfnamefont {A.~J.}\ \bibnamefont
  {Millis}},\ }\bibfield  {title} {\bibinfo {title} {Effect of a nonzero
  temperature on quantum critical points in itinerant fermion systems},\ }\href
  {https://doi.org/10.1103/PhysRevB.48.7183} {\bibfield  {journal} {\bibinfo
  {journal} {Physical Review B}\ }\textbf {\bibinfo {volume} {48}},\ \bibinfo
  {pages} {7183} (\bibinfo {year} {1993})}\BibitemShut {NoStop}%
\bibitem [{\citenamefont {Chubukov}\ \emph {et~al.}(2004)\citenamefont
  {Chubukov}, \citenamefont {P{\'e}pin},\ and\ \citenamefont
  {Rech}}]{chubukovInstabilityQuantumCriticalPoint2004}%
  \BibitemOpen
  \bibfield  {author} {\bibinfo {author} {\bibfnamefont {A.~V.}\ \bibnamefont
  {Chubukov}}, \bibinfo {author} {\bibfnamefont {C.}~\bibnamefont
  {P{\'e}pin}},\ and\ \bibinfo {author} {\bibfnamefont {J.}~\bibnamefont
  {Rech}},\ }\bibfield  {title} {\bibinfo {title} {Instability of the
  quantum-critical point of itinerant ferromagnets},\ }\href
  {https://doi.org/10.1103/PhysRevLett.92.147003} {\bibfield  {journal}
  {\bibinfo  {journal} {Physical Review Letters}\ }\textbf {\bibinfo {volume}
  {92}},\ \bibinfo {pages} {147003} (\bibinfo {year} {2004})}\BibitemShut
  {NoStop}%
\bibitem [{\citenamefont {Metlitski}\ and\ \citenamefont
  {Sachdev}(2010)}]{Metlitski2010a}%
  \BibitemOpen
  \bibfield  {author} {\bibinfo {author} {\bibfnamefont {M.~A.}\ \bibnamefont
  {Metlitski}}\ and\ \bibinfo {author} {\bibfnamefont {S.}~\bibnamefont
  {Sachdev}},\ }\bibfield  {title} {\bibinfo {title} {{Quantum phase
  transitions of metals in two spatial dimensions. I. Ising-nematic order}},\
  }\href {https://doi.org/10.1103/PhysRevB.82.075127} {\bibfield  {journal}
  {\bibinfo  {journal} {Physical Review B - Condensed Matter and Materials
  Physics}\ }\textbf {\bibinfo {volume} {82}},\ \bibinfo {pages} {41} (\bibinfo
  {year} {2010})},\ \Eprint {https://arxiv.org/abs/1001.1153} {arXiv:1001.1153}
  \BibitemShut {NoStop}%
\bibitem [{\citenamefont {Lee}(2018)}]{leeRecentDevelopmentsNonFermi2018}%
  \BibitemOpen
  \bibfield  {author} {\bibinfo {author} {\bibfnamefont {S.-S.}\ \bibnamefont
  {Lee}},\ }\bibfield  {title} {\bibinfo {title} {Recent {{Developments}} in
  {{Non-Fermi Liquid Theory}}},\ }\href@noop {} {\bibfield  {journal} {\bibinfo
   {journal} {Annual Review of Condensed Matter Physics}\ ,\ \bibinfo {pages}
  {20}} (\bibinfo {year} {2018})}\BibitemShut {NoStop}%
\bibitem [{\citenamefont {Damia}\ \emph {et~al.}(2019)\citenamefont {Damia},
  \citenamefont {Kachru}, \citenamefont {Raghu},\ and\ \citenamefont
  {Torroba}}]{Raghu_PRL2019}%
  \BibitemOpen
  \bibfield  {author} {\bibinfo {author} {\bibfnamefont {J.~A.}\ \bibnamefont
  {Damia}}, \bibinfo {author} {\bibfnamefont {S.}~\bibnamefont {Kachru}},
  \bibinfo {author} {\bibfnamefont {S.}~\bibnamefont {Raghu}},\ and\ \bibinfo
  {author} {\bibfnamefont {G.}~\bibnamefont {Torroba}},\ }\bibfield  {title}
  {\bibinfo {title} {Two-dimensional non-fermi-liquid metals: A solvable large-
  {N} limit},\ }\bibfield  {journal} {\bibinfo  {journal} {Physical Review
  Letters}\ }\textbf {\bibinfo {volume} {123}},\ \href
  {https://doi.org/10.1103/physrevlett.123.096402}
  {10.1103/physrevlett.123.096402} (\bibinfo {year} {2019})\BibitemShut
  {NoStop}%
\bibitem [{\citenamefont {Berg}\ \emph {et~al.}(2019)\citenamefont {Berg},
  \citenamefont {Lederer}, \citenamefont {Schattner},\ and\ \citenamefont
  {Trebst}}]{bergMonteCarloStudies2019}%
  \BibitemOpen
  \bibfield  {author} {\bibinfo {author} {\bibfnamefont {E.}~\bibnamefont
  {Berg}}, \bibinfo {author} {\bibfnamefont {S.}~\bibnamefont {Lederer}},
  \bibinfo {author} {\bibfnamefont {Y.}~\bibnamefont {Schattner}},\ and\
  \bibinfo {author} {\bibfnamefont {S.}~\bibnamefont {Trebst}},\ }\bibfield
  {title} {\bibinfo {title} {Monte {{Carlo Studies}} of {{Quantum Critical
  Metals}}},\ }\href {https://doi.org/10.1146/annurev-conmatphys-031218-013339}
  {\bibfield  {journal} {\bibinfo  {journal} {Annual Review of Condensed Matter
  Physics}\ }\textbf {\bibinfo {volume} {10}},\ \bibinfo {pages} {63} (\bibinfo
  {year} {2019})}\BibitemShut {NoStop}%
\bibitem [{\citenamefont {Varma}(2020)}]{VarmaRMP2020}%
  \BibitemOpen
  \bibfield  {author} {\bibinfo {author} {\bibfnamefont {C.~M.}\ \bibnamefont
  {Varma}},\ }\bibfield  {title} {\bibinfo {title} {Colloquium : Linear in
  temperature resistivity and associated mysteries including high temperature
  superconductivity},\ }\bibfield  {journal} {\bibinfo  {journal} {Reviews of
  Modern Physics}\ }\textbf {\bibinfo {volume} {92}},\ \href
  {https://doi.org/10.1103/revmodphys.92.031001} {10.1103/revmodphys.92.031001}
  (\bibinfo {year} {2020})\BibitemShut {NoStop}%
\bibitem [{\citenamefont {Metlitski}\ \emph {et~al.}(2015)\citenamefont
  {Metlitski}, \citenamefont {Mross}, \citenamefont {Sachdev},\ and\
  \citenamefont {Senthil}}]{Metlitski2015}%
  \BibitemOpen
  \bibfield  {author} {\bibinfo {author} {\bibfnamefont {M.~A.}\ \bibnamefont
  {Metlitski}}, \bibinfo {author} {\bibfnamefont {D.~F.}\ \bibnamefont
  {Mross}}, \bibinfo {author} {\bibfnamefont {S.}~\bibnamefont {Sachdev}},\
  and\ \bibinfo {author} {\bibfnamefont {T.}~\bibnamefont {Senthil}},\
  }\bibfield  {title} {\bibinfo {title} {Cooper pairing in non-fermi liquids},\
  }\href {https://doi.org/10.1103/PhysRevB.91.115111} {\bibfield  {journal}
  {\bibinfo  {journal} {Phys. Rev. B}\ }\textbf {\bibinfo {volume} {91}},\
  \bibinfo {pages} {115111} (\bibinfo {year} {2015})}\BibitemShut {NoStop}%
\bibitem [{\citenamefont {Mandal}\ and\ \citenamefont
  {Lee}(2015)}]{Mandal2015}%
  \BibitemOpen
  \bibfield  {author} {\bibinfo {author} {\bibfnamefont {I.}~\bibnamefont
  {Mandal}}\ and\ \bibinfo {author} {\bibfnamefont {S.-S.}\ \bibnamefont
  {Lee}},\ }\bibfield  {title} {\bibinfo {title} {Ultraviolet/infrared mixing
  in non-fermi liquids},\ }\href {https://doi.org/10.1103/PhysRevB.92.035141}
  {\bibfield  {journal} {\bibinfo  {journal} {Phys. Rev. B}\ }\textbf {\bibinfo
  {volume} {92}},\ \bibinfo {pages} {035141} (\bibinfo {year}
  {2015})}\BibitemShut {NoStop}%
\bibitem [{\citenamefont {Mandal}(2016)}]{Mandal2016}%
  \BibitemOpen
  \bibfield  {author} {\bibinfo {author} {\bibfnamefont {I.}~\bibnamefont
  {Mandal}},\ }\bibfield  {title} {\bibinfo {title} {Superconducting
  instability in non-fermi liquids},\ }\href
  {https://doi.org/10.1103/PhysRevB.94.115138} {\bibfield  {journal} {\bibinfo
  {journal} {Phys. Rev. B}\ }\textbf {\bibinfo {volume} {94}},\ \bibinfo
  {pages} {115138} (\bibinfo {year} {2016})}\BibitemShut {NoStop}%
\bibitem [{\citenamefont {Bi}\ \emph {et~al.}(2017)\citenamefont {Bi},
  \citenamefont {Jian}, \citenamefont {You}, \citenamefont {Pawlak},\ and\
  \citenamefont {Xu}}]{ZhenInstability2017}%
  \BibitemOpen
  \bibfield  {author} {\bibinfo {author} {\bibfnamefont {Z.}~\bibnamefont
  {Bi}}, \bibinfo {author} {\bibfnamefont {C.-M.}\ \bibnamefont {Jian}},
  \bibinfo {author} {\bibfnamefont {Y.-Z.}\ \bibnamefont {You}}, \bibinfo
  {author} {\bibfnamefont {K.~A.}\ \bibnamefont {Pawlak}},\ and\ \bibinfo
  {author} {\bibfnamefont {C.}~\bibnamefont {Xu}},\ }\bibfield  {title}
  {\bibinfo {title} {Instability of the non-fermi-liquid state of the
  sachdev-ye-kitaev model},\ }\href
  {https://doi.org/10.1103/PhysRevB.95.205105} {\bibfield  {journal} {\bibinfo
  {journal} {Phys. Rev. B}\ }\textbf {\bibinfo {volume} {95}},\ \bibinfo
  {pages} {205105} (\bibinfo {year} {2017})}\BibitemShut {NoStop}%
\bibitem [{\citenamefont {Cao}\ \emph {et~al.}(2014)\citenamefont {Cao},
  \citenamefont {Rodan-Legrain}, \citenamefont {Park}, \citenamefont {Yuan},
  \citenamefont {Watanabe}, \citenamefont {Taniguchi}, \citenamefont
  {Fernandes}, \citenamefont {Fu},\ and\ \citenamefont
  {Jarillo-Herrero}}]{cao_nematicity_2021}%
  \BibitemOpen
  \bibfield  {author} {\bibinfo {author} {\bibfnamefont {Y.}~\bibnamefont
  {Cao}}, \bibinfo {author} {\bibfnamefont {D.}~\bibnamefont {Rodan-Legrain}},
  \bibinfo {author} {\bibfnamefont {J.~M.}\ \bibnamefont {Park}}, \bibinfo
  {author} {\bibfnamefont {N.~F.~Q.}\ \bibnamefont {Yuan}}, \bibinfo {author}
  {\bibfnamefont {K.}~\bibnamefont {Watanabe}}, \bibinfo {author}
  {\bibfnamefont {T.}~\bibnamefont {Taniguchi}}, \bibinfo {author}
  {\bibfnamefont {R.~M.}\ \bibnamefont {Fernandes}}, \bibinfo {author}
  {\bibfnamefont {L.}~\bibnamefont {Fu}},\ and\ \bibinfo {author}
  {\bibfnamefont {P.}~\bibnamefont {Jarillo-Herrero}},\ }\bibfield  {title}
  {\bibinfo {title} {Nematicity and competing orders in superconducting
  magic-angle graphene},\ }\href {https://doi.org/10.1126/science.abc2836}
  {\bibfield  {journal} {\bibinfo  {journal} {Science}\ }\textbf {\bibinfo
  {volume} {372}},\ \bibinfo {pages} {264} (\bibinfo {year}
  {2014})}\BibitemShut {NoStop}%
\bibitem [{\citenamefont {Cao}\ \emph {et~al.}(2020)\citenamefont {Cao},
  \citenamefont {Chowdhury}, \citenamefont {Rodan-Legrain}, \citenamefont
  {Rubies-Bigorda}, \citenamefont {Watanabe}, \citenamefont {Taniguchi},
  \citenamefont {Senthil},\ and\ \citenamefont
  {Jarillo-Herrero}}]{cao_strange_2020}%
  \BibitemOpen
  \bibfield  {author} {\bibinfo {author} {\bibfnamefont {Y.}~\bibnamefont
  {Cao}}, \bibinfo {author} {\bibfnamefont {D.}~\bibnamefont {Chowdhury}},
  \bibinfo {author} {\bibfnamefont {D.}~\bibnamefont {Rodan-Legrain}}, \bibinfo
  {author} {\bibfnamefont {O.}~\bibnamefont {Rubies-Bigorda}}, \bibinfo
  {author} {\bibfnamefont {K.}~\bibnamefont {Watanabe}}, \bibinfo {author}
  {\bibfnamefont {T.}~\bibnamefont {Taniguchi}}, \bibinfo {author}
  {\bibfnamefont {T.}~\bibnamefont {Senthil}},\ and\ \bibinfo {author}
  {\bibfnamefont {P.}~\bibnamefont {Jarillo-Herrero}},\ }\bibfield  {title}
  {\bibinfo {title} {Strange metal in magic-angle graphene with near planckian
  dissipation},\ }\href {https://doi.org/10.1103/PhysRevLett.124.076801}
  {\bibfield  {journal} {\bibinfo  {journal} {Physical Review Letters}\
  }\textbf {\bibinfo {volume} {124}},\ \bibinfo {pages} {076801} (\bibinfo
  {year} {2020})}\BibitemShut {NoStop}%
\bibitem [{\citenamefont {Rubio-Verdú}\ \emph {et~al.}(2022)\citenamefont
  {Rubio-Verdú}, \citenamefont {Turkel}, \citenamefont {Song}, \citenamefont
  {Klebl}, \citenamefont {Samajdar}, \citenamefont {Scheurer}, \citenamefont
  {Venderbos}, \citenamefont {Watanabe}, \citenamefont {Taniguchi},
  \citenamefont {Ochoa}, \citenamefont {Xian}, \citenamefont {Kennes},
  \citenamefont {Fernandes}, \citenamefont {Rubio},\ and\ \citenamefont
  {Pasupathy}}]{rubio-verdu_moire_2022}%
  \BibitemOpen
  \bibfield  {author} {\bibinfo {author} {\bibfnamefont {C.}~\bibnamefont
  {Rubio-Verdú}}, \bibinfo {author} {\bibfnamefont {S.}~\bibnamefont
  {Turkel}}, \bibinfo {author} {\bibfnamefont {Y.}~\bibnamefont {Song}},
  \bibinfo {author} {\bibfnamefont {L.}~\bibnamefont {Klebl}}, \bibinfo
  {author} {\bibfnamefont {R.}~\bibnamefont {Samajdar}}, \bibinfo {author}
  {\bibfnamefont {M.~S.}\ \bibnamefont {Scheurer}}, \bibinfo {author}
  {\bibfnamefont {J.~W.~F.}\ \bibnamefont {Venderbos}}, \bibinfo {author}
  {\bibfnamefont {K.}~\bibnamefont {Watanabe}}, \bibinfo {author}
  {\bibfnamefont {T.}~\bibnamefont {Taniguchi}}, \bibinfo {author}
  {\bibfnamefont {H.}~\bibnamefont {Ochoa}}, \bibinfo {author} {\bibfnamefont
  {L.}~\bibnamefont {Xian}}, \bibinfo {author} {\bibfnamefont {D.~M.}\
  \bibnamefont {Kennes}}, \bibinfo {author} {\bibfnamefont {R.~M.}\
  \bibnamefont {Fernandes}}, \bibinfo {author} {\bibfnamefont {Ã.}~\bibnamefont
  {Rubio}},\ and\ \bibinfo {author} {\bibfnamefont {A.~N.}\ \bibnamefont
  {Pasupathy}},\ }\bibfield  {title} {\bibinfo {title} {Moir\'{e} nematic phase
  in twisted double bilayer graphene},\ }\href
  {https://doi.org/10.1038/s41567-021-01438-2} {\bibfield  {journal} {\bibinfo
  {journal} {Nature Physics}\ }\textbf {\bibinfo {volume} {18}},\ \bibinfo
  {pages} {196} (\bibinfo {year} {2022})}\BibitemShut {NoStop}%
\bibitem [{\citenamefont {Avci}\ \emph {et~al.}(2014)\citenamefont {Avci},
  \citenamefont {Chmaissem}, \citenamefont {Allred}, \citenamefont
  {Rosenkranz}, \citenamefont {Eremin}, \citenamefont {Chubukov}, \citenamefont
  {Bugaris}, \citenamefont {Chung}, \citenamefont {Kanatzidis}, \citenamefont
  {Castellan}, \citenamefont {Schlueter}, \citenamefont {Claus}, \citenamefont
  {Khalyavin}, \citenamefont {Manuel}, \citenamefont {Daoud-Aladine},\ and\
  \citenamefont {Osborn}}]{avci_magnetically_2014}%
  \BibitemOpen
  \bibfield  {author} {\bibinfo {author} {\bibfnamefont {S.}~\bibnamefont
  {Avci}}, \bibinfo {author} {\bibfnamefont {O.}~\bibnamefont {Chmaissem}},
  \bibinfo {author} {\bibfnamefont {J.}~\bibnamefont {Allred}}, \bibinfo
  {author} {\bibfnamefont {S.}~\bibnamefont {Rosenkranz}}, \bibinfo {author}
  {\bibfnamefont {I.}~\bibnamefont {Eremin}}, \bibinfo {author} {\bibfnamefont
  {A.}~\bibnamefont {Chubukov}}, \bibinfo {author} {\bibfnamefont
  {D.}~\bibnamefont {Bugaris}}, \bibinfo {author} {\bibfnamefont
  {D.}~\bibnamefont {Chung}}, \bibinfo {author} {\bibfnamefont
  {M.}~\bibnamefont {Kanatzidis}}, \bibinfo {author} {\bibfnamefont {J.-P.}\
  \bibnamefont {Castellan}}, \bibinfo {author} {\bibfnamefont {J.}~\bibnamefont
  {Schlueter}}, \bibinfo {author} {\bibfnamefont {H.}~\bibnamefont {Claus}},
  \bibinfo {author} {\bibfnamefont {D.}~\bibnamefont {Khalyavin}}, \bibinfo
  {author} {\bibfnamefont {P.}~\bibnamefont {Manuel}}, \bibinfo {author}
  {\bibfnamefont {A.}~\bibnamefont {Daoud-Aladine}},\ and\ \bibinfo {author}
  {\bibfnamefont {R.}~\bibnamefont {Osborn}},\ }\bibfield  {title} {\bibinfo
  {title} {Magnetically driven suppression of nematic order in an iron-based
  superconductor},\ }\href {https://doi.org/10.1038/ncomms4845} {\bibfield
  {journal} {\bibinfo  {journal} {Nature Communications}\ }\textbf {\bibinfo
  {volume} {5}},\ \bibinfo {pages} {3845} (\bibinfo {year} {2014})}\BibitemShut
  {NoStop}%
\bibitem [{\citenamefont {Fernandes}\ \emph {et~al.}(2014)\citenamefont
  {Fernandes}, \citenamefont {Chubukov},\ and\ \citenamefont
  {Schmalian}}]{fernandesWhatDrivesNematic2014}%
  \BibitemOpen
  \bibfield  {author} {\bibinfo {author} {\bibfnamefont {R.~M.}\ \bibnamefont
  {Fernandes}}, \bibinfo {author} {\bibfnamefont {A.~V.}\ \bibnamefont
  {Chubukov}},\ and\ \bibinfo {author} {\bibfnamefont {J.}~\bibnamefont
  {Schmalian}},\ }\bibfield  {title} {\bibinfo {title} {What drives nematic
  order in iron-based superconductors?},\ }\href
  {https://doi.org/10.1038/nphys2877} {\bibfield  {journal} {\bibinfo
  {journal} {Nature Physics}\ }\textbf {\bibinfo {volume} {10}},\ \bibinfo
  {pages} {97} (\bibinfo {year} {2014})}\BibitemShut {NoStop}%
\bibitem [{\citenamefont {Khasanov}\ \emph {et~al.}(2018)\citenamefont
  {Khasanov}, \citenamefont {Fernandes}, \citenamefont {Simutis}, \citenamefont
  {Guguchia}, \citenamefont {Amato}, \citenamefont {Luetkens}, \citenamefont
  {Morenzoni}, \citenamefont {Dong}, \citenamefont {Zhou},\ and\ \citenamefont
  {Zhao}}]{khasanov_magnetic_2018}%
  \BibitemOpen
  \bibfield  {author} {\bibinfo {author} {\bibfnamefont {R.}~\bibnamefont
  {Khasanov}}, \bibinfo {author} {\bibfnamefont {R.~M.}\ \bibnamefont
  {Fernandes}}, \bibinfo {author} {\bibfnamefont {G.}~\bibnamefont {Simutis}},
  \bibinfo {author} {\bibfnamefont {Z.}~\bibnamefont {Guguchia}}, \bibinfo
  {author} {\bibfnamefont {A.}~\bibnamefont {Amato}}, \bibinfo {author}
  {\bibfnamefont {H.}~\bibnamefont {Luetkens}}, \bibinfo {author}
  {\bibfnamefont {E.}~\bibnamefont {Morenzoni}}, \bibinfo {author}
  {\bibfnamefont {X.}~\bibnamefont {Dong}}, \bibinfo {author} {\bibfnamefont
  {F.}~\bibnamefont {Zhou}},\ and\ \bibinfo {author} {\bibfnamefont
  {Z.}~\bibnamefont {Zhao}},\ }\bibfield  {title} {\bibinfo {title} {Magnetic
  tricritical point and nematicity in {FeSe} under pressure},\ }\href
  {https://doi.org/10.1103/PhysRevB.97.224510} {\bibfield  {journal} {\bibinfo
  {journal} {Physical Review B}\ }\textbf {\bibinfo {volume} {97}},\ \bibinfo
  {pages} {224510} (\bibinfo {year} {2018})}\BibitemShut {NoStop}%
\bibitem [{\citenamefont {Licciardello}\ \emph {et~al.}(2019)\citenamefont
  {Licciardello}, \citenamefont {Buhot}, \citenamefont {Lu}, \citenamefont
  {Ayres}, \citenamefont {Kasahara}, \citenamefont {Matsuda}, \citenamefont
  {Shibauchi},\ and\ \citenamefont {Hussey}}]{Licciardello2019a}%
  \BibitemOpen
  \bibfield  {author} {\bibinfo {author} {\bibfnamefont {S.}~\bibnamefont
  {Licciardello}}, \bibinfo {author} {\bibfnamefont {J.}~\bibnamefont {Buhot}},
  \bibinfo {author} {\bibfnamefont {J.}~\bibnamefont {Lu}}, \bibinfo {author}
  {\bibfnamefont {J.}~\bibnamefont {Ayres}}, \bibinfo {author} {\bibfnamefont
  {S.}~\bibnamefont {Kasahara}}, \bibinfo {author} {\bibfnamefont
  {Y.}~\bibnamefont {Matsuda}}, \bibinfo {author} {\bibfnamefont
  {T.}~\bibnamefont {Shibauchi}},\ and\ \bibinfo {author} {\bibfnamefont
  {N.~E.}\ \bibnamefont {Hussey}},\ }\bibfield  {title} {\bibinfo {title}
  {{Electrical resistivity across a nematic quantum critical point}},\ }\href
  {https://doi.org/10.1038/s41586-019-0923-y} {\bibfield  {journal} {\bibinfo
  {journal} {Nature}\ }\textbf {\bibinfo {volume} {567}},\ \bibinfo {pages}
  {213} (\bibinfo {year} {2019})}\BibitemShut {NoStop}%
\bibitem [{\citenamefont {Culo}\ \emph {et~al.}(2021)\citenamefont {Culo},
  \citenamefont {Berben}, \citenamefont {Hsu}, \citenamefont {Ayres},
  \citenamefont {Hinlopen}, \citenamefont {Kasahara}, \citenamefont {Matsuda},
  \citenamefont {Shibauchi},\ and\ \citenamefont
  {Hussey}}]{culo_putative_2021}%
  \BibitemOpen
  \bibfield  {author} {\bibinfo {author} {\bibfnamefont {M.}~\bibnamefont
  {Culo}}, \bibinfo {author} {\bibfnamefont {M.}~\bibnamefont {Berben}},
  \bibinfo {author} {\bibfnamefont {Y.-T.}\ \bibnamefont {Hsu}}, \bibinfo
  {author} {\bibfnamefont {J.}~\bibnamefont {Ayres}}, \bibinfo {author}
  {\bibfnamefont {R.~D.~H.}\ \bibnamefont {Hinlopen}}, \bibinfo {author}
  {\bibfnamefont {S.}~\bibnamefont {Kasahara}}, \bibinfo {author}
  {\bibfnamefont {Y.}~\bibnamefont {Matsuda}}, \bibinfo {author} {\bibfnamefont
  {T.}~\bibnamefont {Shibauchi}},\ and\ \bibinfo {author} {\bibfnamefont
  {N.~E.}\ \bibnamefont {Hussey}},\ }\bibfield  {title} {\bibinfo {title}
  {Putative {Hall} response of the strange metal component in {Fe
  Se$_{1-x}$S$_x$}},\ }\href {https://doi.org/10.1103/PhysRevResearch.3.023069}
  {\bibfield  {journal} {\bibinfo  {journal} {Physical Review Research}\
  }\textbf {\bibinfo {volume} {3}},\ \bibinfo {pages} {023069} (\bibinfo {year}
  {2021})}\BibitemShut {NoStop}%
\bibitem [{\citenamefont {Steffensen}\ \emph {et~al.}(2021)\citenamefont
  {Steffensen}, \citenamefont {Kreisel}, \citenamefont {Hirschfeld},\ and\
  \citenamefont {Andersen}}]{Hirschfeld2021}%
  \BibitemOpen
  \bibfield  {author} {\bibinfo {author} {\bibfnamefont {D.}~\bibnamefont
  {Steffensen}}, \bibinfo {author} {\bibfnamefont {A.}~\bibnamefont {Kreisel}},
  \bibinfo {author} {\bibfnamefont {P.~J.}\ \bibnamefont {Hirschfeld}},\ and\
  \bibinfo {author} {\bibfnamefont {B.~M.}\ \bibnamefont {Andersen}},\
  }\bibfield  {title} {\bibinfo {title} {Interorbital nematicity and the origin
  of a single electron fermi pocket in {FeSe}},\ }\href
  {https://doi.org/10.1103/physrevb.103.054505} {\bibfield  {journal} {\bibinfo
   {journal} {Physical Review B}\ }\textbf {\bibinfo {volume} {103}},\ \bibinfo
  {pages} {054505} (\bibinfo {year} {2021})}\BibitemShut {NoStop}%
\bibitem [{\citenamefont {Nie}\ \emph {et~al.}(2017)\citenamefont {Nie},
  \citenamefont {Maharaj}, \citenamefont {Fradkin},\ and\ \citenamefont
  {Kivelson}}]{nie_vestigial_2017}%
  \BibitemOpen
  \bibfield  {author} {\bibinfo {author} {\bibfnamefont {L.}~\bibnamefont
  {Nie}}, \bibinfo {author} {\bibfnamefont {A.~V.}\ \bibnamefont {Maharaj}},
  \bibinfo {author} {\bibfnamefont {E.}~\bibnamefont {Fradkin}},\ and\ \bibinfo
  {author} {\bibfnamefont {S.~A.}\ \bibnamefont {Kivelson}},\ }\bibfield
  {title} {\bibinfo {title} {Vestigial nematicity from spin and/or charge order
  in the cuprates},\ }\href {https://doi.org/10.1103/PhysRevB.96.085142}
  {\bibfield  {journal} {\bibinfo  {journal} {Physical Review B}\ }\textbf
  {\bibinfo {volume} {96}},\ \bibinfo {pages} {085142} (\bibinfo {year}
  {2017})}\BibitemShut {NoStop}%
\bibitem [{\citenamefont {Mukhopadhyay}\ \emph {et~al.}(2019)\citenamefont
  {Mukhopadhyay}, \citenamefont {Sharma}, \citenamefont {Kim}, \citenamefont
  {Edkins}, \citenamefont {Hamidian}, \citenamefont {Eisaki}, \citenamefont
  {Uchida}, \citenamefont {Kim}, \citenamefont {Lawler}, \citenamefont
  {Mackenzie}, \citenamefont {Davis},\ and\ \citenamefont
  {Fujita}}]{mukhopadhyay_evidence_2019}%
  \BibitemOpen
  \bibfield  {author} {\bibinfo {author} {\bibfnamefont {S.}~\bibnamefont
  {Mukhopadhyay}}, \bibinfo {author} {\bibfnamefont {R.}~\bibnamefont
  {Sharma}}, \bibinfo {author} {\bibfnamefont {C.~K.}\ \bibnamefont {Kim}},
  \bibinfo {author} {\bibfnamefont {S.~D.}\ \bibnamefont {Edkins}}, \bibinfo
  {author} {\bibfnamefont {M.~H.}\ \bibnamefont {Hamidian}}, \bibinfo {author}
  {\bibfnamefont {H.}~\bibnamefont {Eisaki}}, \bibinfo {author} {\bibfnamefont
  {S.-i.}\ \bibnamefont {Uchida}}, \bibinfo {author} {\bibfnamefont {E.-A.}\
  \bibnamefont {Kim}}, \bibinfo {author} {\bibfnamefont {M.~J.}\ \bibnamefont
  {Lawler}}, \bibinfo {author} {\bibfnamefont {A.~P.}\ \bibnamefont
  {Mackenzie}}, \bibinfo {author} {\bibfnamefont {J.~C.~S.}\ \bibnamefont
  {Davis}},\ and\ \bibinfo {author} {\bibfnamefont {K.}~\bibnamefont
  {Fujita}},\ }\bibfield  {title} {\bibinfo {title} {Evidence for a vestigial
  nematic state in the cuprate pseudogap phase},\ }\href
  {https://doi.org/10.1073/pnas.1821454116} {\bibfield  {journal} {\bibinfo
  {journal} {Proceedings of the National Academy of Sciences}\ }\textbf
  {\bibinfo {volume} {116}},\ \bibinfo {pages} {13249} (\bibinfo {year}
  {2019})}\BibitemShut {NoStop}%
\bibitem [{\citenamefont {Gupta}\ \emph {et~al.}(2021)\citenamefont {Gupta},
  \citenamefont {{McMahon}}, \citenamefont {Sutarto}, \citenamefont {Shi},
  \citenamefont {Gong}, \citenamefont {Wei}, \citenamefont {Shen},
  \citenamefont {He}, \citenamefont {Ma}, \citenamefont {Dragomir},
  \citenamefont {Gaulin},\ and\ \citenamefont
  {Hawthorn}}]{gupta_vanishing_2021}%
  \BibitemOpen
  \bibfield  {author} {\bibinfo {author} {\bibfnamefont {N.~K.}\ \bibnamefont
  {Gupta}}, \bibinfo {author} {\bibfnamefont {C.}~\bibnamefont {{McMahon}}},
  \bibinfo {author} {\bibfnamefont {R.}~\bibnamefont {Sutarto}}, \bibinfo
  {author} {\bibfnamefont {T.}~\bibnamefont {Shi}}, \bibinfo {author}
  {\bibfnamefont {R.}~\bibnamefont {Gong}}, \bibinfo {author} {\bibfnamefont
  {H.~I.}\ \bibnamefont {Wei}}, \bibinfo {author} {\bibfnamefont {K.~M.}\
  \bibnamefont {Shen}}, \bibinfo {author} {\bibfnamefont {F.}~\bibnamefont
  {He}}, \bibinfo {author} {\bibfnamefont {Q.}~\bibnamefont {Ma}}, \bibinfo
  {author} {\bibfnamefont {M.}~\bibnamefont {Dragomir}}, \bibinfo {author}
  {\bibfnamefont {B.~D.}\ \bibnamefont {Gaulin}},\ and\ \bibinfo {author}
  {\bibfnamefont {D.~G.}\ \bibnamefont {Hawthorn}},\ }\bibfield  {title}
  {\bibinfo {title} {Vanishing nematic order beyond the pseudogap phase in
  overdoped cuprate superconductors},\ }\href
  {https://doi.org/10.1073/pnas.2106881118} {\bibfield  {journal} {\bibinfo
  {journal} {Proceedings of the National Academy of Sciences}\ }\textbf
  {\bibinfo {volume} {118}},\ \bibinfo {pages} {e2106881118} (\bibinfo {year}
  {2021})}\BibitemShut {NoStop}%
\bibitem [{\citenamefont {Borzi}\ \emph {et~al.}(2007)\citenamefont {Borzi},
  \citenamefont {Grigera}, \citenamefont {Farrell}, \citenamefont {Perry},
  \citenamefont {Lister}, \citenamefont {Lee}, \citenamefont {Tennant},
  \citenamefont {Maeno},\ and\ \citenamefont
  {Mackenzie}}]{Mackenzie_Science2007}%
  \BibitemOpen
  \bibfield  {author} {\bibinfo {author} {\bibfnamefont {R.~A.}\ \bibnamefont
  {Borzi}}, \bibinfo {author} {\bibfnamefont {S.~A.}\ \bibnamefont {Grigera}},
  \bibinfo {author} {\bibfnamefont {J.}~\bibnamefont {Farrell}}, \bibinfo
  {author} {\bibfnamefont {R.~S.}\ \bibnamefont {Perry}}, \bibinfo {author}
  {\bibfnamefont {S.~J.~S.}\ \bibnamefont {Lister}}, \bibinfo {author}
  {\bibfnamefont {S.~L.}\ \bibnamefont {Lee}}, \bibinfo {author} {\bibfnamefont
  {D.~A.}\ \bibnamefont {Tennant}}, \bibinfo {author} {\bibfnamefont
  {Y.}~\bibnamefont {Maeno}},\ and\ \bibinfo {author} {\bibfnamefont {A.~P.}\
  \bibnamefont {Mackenzie}},\ }\bibfield  {title} {\bibinfo {title} {Formation
  of a nematic fluid at high fields in {Sr$_3$Ru$_2$O$_7$}},\ }\href
  {https://doi.org/10.1126/science.1134796} {\bibfield  {journal} {\bibinfo
  {journal} {Science}\ }\textbf {\bibinfo {volume} {315}},\ \bibinfo {pages}
  {214–217} (\bibinfo {year} {2007})}\BibitemShut {NoStop}%
\bibitem [{\citenamefont {Kee}\ and\ \citenamefont {Kim}(2005)}]{kee2005}%
  \BibitemOpen
  \bibfield  {author} {\bibinfo {author} {\bibfnamefont {H.-Y.}\ \bibnamefont
  {Kee}}\ and\ \bibinfo {author} {\bibfnamefont {Y.~B.}\ \bibnamefont {Kim}},\
  }\bibfield  {title} {\bibinfo {title} {Itinerant metamagnetism induced by
  electronic nematic order},\ }\href
  {https://doi.org/10.1103/physrevb.71.184402} {\bibfield  {journal} {\bibinfo
  {journal} {Physical Review B}\ }\textbf {\bibinfo {volume} {71}},\ \bibinfo
  {pages} {184402} (\bibinfo {year} {2005})}\BibitemShut {NoStop}%
\bibitem [{\citenamefont {Raghu}\ \emph {et~al.}(2009)\citenamefont {Raghu},
  \citenamefont {Paramekanti}, \citenamefont {Kim}, \citenamefont {Borzi},
  \citenamefont {Grigera}, \citenamefont {Mackenzie},\ and\ \citenamefont
  {Kivelson}}]{Raghu_PRB2009}%
  \BibitemOpen
  \bibfield  {author} {\bibinfo {author} {\bibfnamefont {S.}~\bibnamefont
  {Raghu}}, \bibinfo {author} {\bibfnamefont {A.}~\bibnamefont {Paramekanti}},
  \bibinfo {author} {\bibfnamefont {E.~A.}\ \bibnamefont {Kim}}, \bibinfo
  {author} {\bibfnamefont {R.~A.}\ \bibnamefont {Borzi}}, \bibinfo {author}
  {\bibfnamefont {S.~A.}\ \bibnamefont {Grigera}}, \bibinfo {author}
  {\bibfnamefont {A.~P.}\ \bibnamefont {Mackenzie}},\ and\ \bibinfo {author}
  {\bibfnamefont {S.~A.}\ \bibnamefont {Kivelson}},\ }\bibfield  {title}
  {\bibinfo {title} {Microscopic theory of the nematic phase in
  {Sr$_3$Ru$_2$O$_7$}},\ }\bibfield  {journal} {\bibinfo  {journal} {Physical
  Review B}\ }\textbf {\bibinfo {volume} {79}},\ \href
  {https://doi.org/10.1103/physrevb.79.214402} {10.1103/physrevb.79.214402}
  (\bibinfo {year} {2009})\BibitemShut {NoStop}%
\bibitem [{\citenamefont {Lester}\ \emph {et~al.}(2015)\citenamefont {Lester},
  \citenamefont {Ramos}, \citenamefont {Perry}, \citenamefont {Croft},
  \citenamefont {Bewley}, \citenamefont {Guidi}, \citenamefont {Manuel},
  \citenamefont {Khalyavin}, \citenamefont {Forgan},\ and\ \citenamefont
  {Hayden}}]{Hayden_NatMat2015}%
  \BibitemOpen
  \bibfield  {author} {\bibinfo {author} {\bibfnamefont {C.}~\bibnamefont
  {Lester}}, \bibinfo {author} {\bibfnamefont {S.}~\bibnamefont {Ramos}},
  \bibinfo {author} {\bibfnamefont {R.~S.}\ \bibnamefont {Perry}}, \bibinfo
  {author} {\bibfnamefont {T.~P.}\ \bibnamefont {Croft}}, \bibinfo {author}
  {\bibfnamefont {R.~I.}\ \bibnamefont {Bewley}}, \bibinfo {author}
  {\bibfnamefont {T.}~\bibnamefont {Guidi}}, \bibinfo {author} {\bibfnamefont
  {P.}~\bibnamefont {Manuel}}, \bibinfo {author} {\bibfnamefont {D.~D.}\
  \bibnamefont {Khalyavin}}, \bibinfo {author} {\bibfnamefont {E.}~\bibnamefont
  {Forgan}},\ and\ \bibinfo {author} {\bibfnamefont {S.~M.}\ \bibnamefont
  {Hayden}},\ }\bibfield  {title} {\bibinfo {title} {Field-tunable
  spin-density-wave phases in {Sr$_3$Ru$_2$O$_7$}},\ }\href
  {https://doi.org/10.1038/nmat4181} {\bibfield  {journal} {\bibinfo  {journal}
  {Nature Materials}\ }\textbf {\bibinfo {volume} {14}},\ \bibinfo {pages}
  {373–378} (\bibinfo {year} {2015})}\BibitemShut {NoStop}%
\bibitem [{\citenamefont {Cooper}\ \emph {et~al.}(2002)\citenamefont {Cooper},
  \citenamefont {Lilly}, \citenamefont {Eisenstein}, \citenamefont {Pfeiffer},\
  and\ \citenamefont {West}}]{Eisenstein_Nematic2002}%
  \BibitemOpen
  \bibfield  {author} {\bibinfo {author} {\bibfnamefont {K.~B.}\ \bibnamefont
  {Cooper}}, \bibinfo {author} {\bibfnamefont {M.~P.}\ \bibnamefont {Lilly}},
  \bibinfo {author} {\bibfnamefont {J.~P.}\ \bibnamefont {Eisenstein}},
  \bibinfo {author} {\bibfnamefont {L.~N.}\ \bibnamefont {Pfeiffer}},\ and\
  \bibinfo {author} {\bibfnamefont {K.~W.}\ \bibnamefont {West}},\ }\bibfield
  {title} {\bibinfo {title} {Onset of anisotropic transport of two-dimensional
  electrons in high landau levels: Possible isotropic-to-nematic liquid-crystal
  phase transition},\ }\href {https://doi.org/10.1103/physrevb.65.241313}
  {\bibfield  {journal} {\bibinfo  {journal} {Physical Review B}\ }\textbf
  {\bibinfo {volume} {65}},\ \bibinfo {pages} {65.241313} (\bibinfo {year}
  {2002})}\BibitemShut {NoStop}%
\bibitem [{\citenamefont {Fradkin}\ \emph {et~al.}(2010)\citenamefont
  {Fradkin}, \citenamefont {Kivelson}, \citenamefont {Lawler}, \citenamefont
  {Eisenstein},\ and\ \citenamefont {Mackenzie}}]{Fradkin2010a}%
  \BibitemOpen
  \bibfield  {author} {\bibinfo {author} {\bibfnamefont {E.}~\bibnamefont
  {Fradkin}}, \bibinfo {author} {\bibfnamefont {S.~A.}\ \bibnamefont
  {Kivelson}}, \bibinfo {author} {\bibfnamefont {M.~J.}\ \bibnamefont
  {Lawler}}, \bibinfo {author} {\bibfnamefont {J.~P.}\ \bibnamefont
  {Eisenstein}},\ and\ \bibinfo {author} {\bibfnamefont {A.~P.}\ \bibnamefont
  {Mackenzie}},\ }\bibfield  {title} {\bibinfo {title} {{Nematic fermi fluids
  in condensed matter physics}},\ }\href
  {https://doi.org/10.1146/annurev-conmatphys-070909-103925} {\bibfield
  {journal} {\bibinfo  {journal} {Annual Review of Condensed Matter Physics}\
  }\textbf {\bibinfo {volume} {1}},\ \bibinfo {pages} {153} (\bibinfo {year}
  {2010})},\ \Eprint {https://arxiv.org/abs/0910.4166} {arXiv:0910.4166}
  \BibitemShut {NoStop}%
\bibitem [{\citenamefont {Xia}\ \emph {et~al.}(2011)\citenamefont {Xia},
  \citenamefont {Eisenstein}, \citenamefont {Pfeiffer},\ and\ \citenamefont
  {West}}]{Xia2011a}%
  \BibitemOpen
  \bibfield  {author} {\bibinfo {author} {\bibfnamefont {J.}~\bibnamefont
  {Xia}}, \bibinfo {author} {\bibfnamefont {J.~P.}\ \bibnamefont {Eisenstein}},
  \bibinfo {author} {\bibfnamefont {L.~N.}\ \bibnamefont {Pfeiffer}},\ and\
  \bibinfo {author} {\bibfnamefont {K.~W.}\ \bibnamefont {West}},\ }\bibfield
  {title} {\bibinfo {title} {{Evidence for a fractionally quantized Hall state
  with anisotropic longitudinal transport}},\ }\href
  {https://doi.org/10.1038/nphys2118} {\bibfield  {journal} {\bibinfo
  {journal} {Nature Physics}\ }\textbf {\bibinfo {volume} {7}},\ \bibinfo
  {pages} {845} (\bibinfo {year} {2011})}\BibitemShut {NoStop}%
\bibitem [{\citenamefont {Feldman}\ \emph {et~al.}(2016)\citenamefont
  {Feldman}, \citenamefont {Randeria}, \citenamefont {Gyenis}, \citenamefont
  {Wu}, \citenamefont {Ji}, \citenamefont {Cava}, \citenamefont {MacDonald},\
  and\ \citenamefont {Yazdani}}]{Feldman2016}%
  \BibitemOpen
  \bibfield  {author} {\bibinfo {author} {\bibfnamefont {B.~E.}\ \bibnamefont
  {Feldman}}, \bibinfo {author} {\bibfnamefont {M.~T.}\ \bibnamefont
  {Randeria}}, \bibinfo {author} {\bibfnamefont {A.}~\bibnamefont {Gyenis}},
  \bibinfo {author} {\bibfnamefont {F.}~\bibnamefont {Wu}}, \bibinfo {author}
  {\bibfnamefont {H.}~\bibnamefont {Ji}}, \bibinfo {author} {\bibfnamefont
  {R.~J.}\ \bibnamefont {Cava}}, \bibinfo {author} {\bibfnamefont {A.~H.}\
  \bibnamefont {MacDonald}},\ and\ \bibinfo {author} {\bibfnamefont
  {A.}~\bibnamefont {Yazdani}},\ }\bibfield  {title} {\bibinfo {title}
  {Observation of a nematic quantum {Hall} liquid on the surface of
  {Bismuth}},\ }\href@noop {} {\bibfield  {journal} {\bibinfo  {journal}
  {Science}\ }\textbf {\bibinfo {volume} {354}} (\bibinfo {year}
  {2016})}\BibitemShut {NoStop}%
\bibitem [{\citenamefont {Hayes}\ \emph {et~al.}(2021)\citenamefont {Hayes},
  \citenamefont {Maksimovic}, \citenamefont {Lopez}, \citenamefont {Chan},
  \citenamefont {Ramshaw}, \citenamefont {McDonald},\ and\ \citenamefont
  {Analytis}}]{Hayes2021a}%
  \BibitemOpen
  \bibfield  {author} {\bibinfo {author} {\bibfnamefont {I.~M.}\ \bibnamefont
  {Hayes}}, \bibinfo {author} {\bibfnamefont {N.}~\bibnamefont {Maksimovic}},
  \bibinfo {author} {\bibfnamefont {G.~N.}\ \bibnamefont {Lopez}}, \bibinfo
  {author} {\bibfnamefont {M.~K.}\ \bibnamefont {Chan}}, \bibinfo {author}
  {\bibfnamefont {B.~J.}\ \bibnamefont {Ramshaw}}, \bibinfo {author}
  {\bibfnamefont {R.~D.}\ \bibnamefont {McDonald}},\ and\ \bibinfo {author}
  {\bibfnamefont {J.~G.}\ \bibnamefont {Analytis}},\ }\bibfield  {title}
  {\bibinfo {title} {{Superconductivity and quantum criticality linked by the
  Hall effect in a strange metal}},\ }\href
  {https://doi.org/10.1038/s41567-020-0982-x} {\bibfield  {journal} {\bibinfo
  {journal} {Nature Physics}\ }\textbf {\bibinfo {volume} {17}},\ \bibinfo
  {pages} {58} (\bibinfo {year} {2021})}\BibitemShut {NoStop}%
\bibitem [{\citenamefont {Schattner}\ \emph {et~al.}(2016)\citenamefont
  {Schattner}, \citenamefont {Lederer}, \citenamefont {Kivelson},\ and\
  \citenamefont {Berg}}]{schattner_ising_2016}%
  \BibitemOpen
  \bibfield  {author} {\bibinfo {author} {\bibfnamefont {Y.}~\bibnamefont
  {Schattner}}, \bibinfo {author} {\bibfnamefont {S.}~\bibnamefont {Lederer}},
  \bibinfo {author} {\bibfnamefont {S.~A.}\ \bibnamefont {Kivelson}},\ and\
  \bibinfo {author} {\bibfnamefont {E.}~\bibnamefont {Berg}},\ }\bibfield
  {title} {\bibinfo {title} {Ising nematic quantum critical point in a metal: A
  {Monte Carlo} study},\ }\href {https://doi.org/10.1103/PhysRevX.6.031028}
  {\bibfield  {journal} {\bibinfo  {journal} {Physical Review X}\ }\textbf
  {\bibinfo {volume} {6}},\ \bibinfo {pages} {031028} (\bibinfo {year}
  {2016})}\BibitemShut {NoStop}%
\bibitem [{\citenamefont {Kuo}\ \emph {et~al.}(2013)\citenamefont {Kuo},
  \citenamefont {Shapiro}, \citenamefont {Riggs},\ and\ \citenamefont
  {Fisher}}]{kuoMeasurementElastoresistivityCoefficients2013}%
  \BibitemOpen
  \bibfield  {author} {\bibinfo {author} {\bibfnamefont {H.-H.}\ \bibnamefont
  {Kuo}}, \bibinfo {author} {\bibfnamefont {M.~C.}\ \bibnamefont {Shapiro}},
  \bibinfo {author} {\bibfnamefont {S.~C.}\ \bibnamefont {Riggs}},\ and\
  \bibinfo {author} {\bibfnamefont {I.~R.}\ \bibnamefont {Fisher}},\ }\bibfield
   {title} {\bibinfo {title} {Measurement of the elastoresistivity coefficients
  of the underdoped iron arsenide {{Ba}}({{Fe}}$_{0.975}$
  {{Co}}$_{0.025}$)$_2${{As}}$_2$},\ }\href
  {https://doi.org/10.1103/PhysRevB.88.085113} {\bibfield  {journal} {\bibinfo
  {journal} {Physical Review B}\ }\textbf {\bibinfo {volume} {88}},\ \bibinfo
  {pages} {085113} (\bibinfo {year} {2013})}\BibitemShut {NoStop}%
\bibitem [{\citenamefont {Kuo}\ \emph {et~al.}(2016)\citenamefont {Kuo},
  \citenamefont {Chu}, \citenamefont {Palmstrom}, \citenamefont {Kivelson},\
  and\ \citenamefont {Fisher}}]{kuo_strain_science2016}%
  \BibitemOpen
  \bibfield  {author} {\bibinfo {author} {\bibfnamefont {H.-H.}\ \bibnamefont
  {Kuo}}, \bibinfo {author} {\bibfnamefont {J.-H.}\ \bibnamefont {Chu}},
  \bibinfo {author} {\bibfnamefont {J.~C.}\ \bibnamefont {Palmstrom}}, \bibinfo
  {author} {\bibfnamefont {S.~A.}\ \bibnamefont {Kivelson}},\ and\ \bibinfo
  {author} {\bibfnamefont {I.~R.}\ \bibnamefont {Fisher}},\ }\bibfield  {title}
  {\bibinfo {title} {Ubiquitous signatures of nematic quantum criticality in
  optimally doped fe-based superconductors},\ }\href
  {https://doi.org/10.1126/science.aab0103} {\bibfield  {journal} {\bibinfo
  {journal} {Science}\ }\textbf {\bibinfo {volume} {352}},\ \bibinfo {pages}
  {958} (\bibinfo {year} {2016})},\ \Eprint
  {https://arxiv.org/abs/https://www.science.org/doi/pdf/10.1126/science.aab0103}
  {https://www.science.org/doi/pdf/10.1126/science.aab0103} \BibitemShut
  {NoStop}%
\bibitem [{\citenamefont {Mirri}\ \emph {et~al.}(2016)\citenamefont {Mirri},
  \citenamefont {Dusza}, \citenamefont {Bastelberger}, \citenamefont
  {Chinotti}, \citenamefont {Chu}, \citenamefont {Kuo}, \citenamefont
  {Fisher},\ and\ \citenamefont {Degiorgi}}]{mirri_electrodynamic_2016}%
  \BibitemOpen
  \bibfield  {author} {\bibinfo {author} {\bibfnamefont {C.}~\bibnamefont
  {Mirri}}, \bibinfo {author} {\bibfnamefont {A.}~\bibnamefont {Dusza}},
  \bibinfo {author} {\bibfnamefont {S.}~\bibnamefont {Bastelberger}}, \bibinfo
  {author} {\bibfnamefont {M.}~\bibnamefont {Chinotti}}, \bibinfo {author}
  {\bibfnamefont {J.-H.}\ \bibnamefont {Chu}}, \bibinfo {author} {\bibfnamefont
  {H.-H.}\ \bibnamefont {Kuo}}, \bibinfo {author} {\bibfnamefont {I.~R.}\
  \bibnamefont {Fisher}},\ and\ \bibinfo {author} {\bibfnamefont
  {L.}~\bibnamefont {Degiorgi}},\ }\bibfield  {title} {\bibinfo {title}
  {Electrodynamic response in the electronic nematic phase of
  {BaFe$_2$As$_2$}},\ }\href {https://doi.org/10.1103/PhysRevB.93.085114}
  {\bibfield  {journal} {\bibinfo  {journal} {Physical Review B}\ }\textbf
  {\bibinfo {volume} {93}},\ \bibinfo {pages} {085114} (\bibinfo {year}
  {2016})}\BibitemShut {NoStop}%
\bibitem [{\citenamefont {Palmstrom}\ \emph {et~al.}(2022)\citenamefont
  {Palmstrom}, \citenamefont {Walmsley}, \citenamefont {Straquadine},
  \citenamefont {Sorensen}, \citenamefont {Hannahs}, \citenamefont {Burns},\
  and\ \citenamefont {Fisher}}]{palmstromComparisonTemperatureDoping2022}%
  \BibitemOpen
  \bibfield  {author} {\bibinfo {author} {\bibfnamefont {J.~C.}\ \bibnamefont
  {Palmstrom}}, \bibinfo {author} {\bibfnamefont {P.}~\bibnamefont {Walmsley}},
  \bibinfo {author} {\bibfnamefont {J.~A.~W.}\ \bibnamefont {Straquadine}},
  \bibinfo {author} {\bibfnamefont {M.~E.}\ \bibnamefont {Sorensen}}, \bibinfo
  {author} {\bibfnamefont {S.~T.}\ \bibnamefont {Hannahs}}, \bibinfo {author}
  {\bibfnamefont {D.~H.}\ \bibnamefont {Burns}},\ and\ \bibinfo {author}
  {\bibfnamefont {I.~R.}\ \bibnamefont {Fisher}},\ }\bibfield  {title}
  {\bibinfo {title} {Comparison of temperature and doping dependence of
  elastoresistivity near a putative nematic quantum critical point},\ }\href
  {https://doi.org/10.1038/s41467-022-28583-3} {\bibfield  {journal} {\bibinfo
  {journal} {Nature Communications}\ }\textbf {\bibinfo {volume} {13}},\
  \bibinfo {pages} {1011} (\bibinfo {year} {2022})}\BibitemShut {NoStop}%
\bibitem [{\citenamefont {Bartlett}\ \emph {et~al.}(2021)\citenamefont
  {Bartlett}, \citenamefont {Steppke}, \citenamefont {Hosoi}, \citenamefont
  {Noad}, \citenamefont {Park}, \citenamefont {Timm}, \citenamefont
  {Shibauchi}, \citenamefont {Mackenzie},\ and\ \citenamefont
  {Hicks}}]{Hicks_PRX2021}%
  \BibitemOpen
  \bibfield  {author} {\bibinfo {author} {\bibfnamefont {J.~M.}\ \bibnamefont
  {Bartlett}}, \bibinfo {author} {\bibfnamefont {A.}~\bibnamefont {Steppke}},
  \bibinfo {author} {\bibfnamefont {S.}~\bibnamefont {Hosoi}}, \bibinfo
  {author} {\bibfnamefont {H.}~\bibnamefont {Noad}}, \bibinfo {author}
  {\bibfnamefont {J.}~\bibnamefont {Park}}, \bibinfo {author} {\bibfnamefont
  {C.}~\bibnamefont {Timm}}, \bibinfo {author} {\bibfnamefont {T.}~\bibnamefont
  {Shibauchi}}, \bibinfo {author} {\bibfnamefont {A.~P.}\ \bibnamefont
  {Mackenzie}},\ and\ \bibinfo {author} {\bibfnamefont {C.~W.}\ \bibnamefont
  {Hicks}},\ }\bibfield  {title} {\bibinfo {title} {Relationship between
  transport anisotropy and nematicity in {FeSe}},\ }\bibfield  {journal}
  {\bibinfo  {journal} {Physical Review X}\ }\textbf {\bibinfo {volume} {11}},\
  \href {https://doi.org/10.1103/physrevx.11.021038}
  {10.1103/physrevx.11.021038} (\bibinfo {year} {2021})\BibitemShut {NoStop}%
\bibitem [{\citenamefont {Arciniaga}\ \emph {et~al.}(2020)\citenamefont
  {Arciniaga}, \citenamefont {Mai},\ and\ \citenamefont
  {Shastry}}]{arciniaga_theory_2020}%
  \BibitemOpen
  \bibfield  {author} {\bibinfo {author} {\bibfnamefont {M.}~\bibnamefont
  {Arciniaga}}, \bibinfo {author} {\bibfnamefont {P.}~\bibnamefont {Mai}},\
  and\ \bibinfo {author} {\bibfnamefont {B.~S.}\ \bibnamefont {Shastry}},\
  }\bibfield  {title} {\bibinfo {title} {Theory of anisotropic
  elastoresistivity of two-dimensional extremely strongly correlated metals},\
  }\href {https://doi.org/10.1103/PhysRevB.101.245149} {\bibfield  {journal}
  {\bibinfo  {journal} {Physical Review B}\ }\textbf {\bibinfo {volume}
  {101}},\ \bibinfo {pages} {245149} (\bibinfo {year} {2020})}\BibitemShut
  {NoStop}%
\bibitem [{\citenamefont {Sachdev}\ and\ \citenamefont
  {Ye}(1993)}]{sachdev1993gapless}%
  \BibitemOpen
  \bibfield  {author} {\bibinfo {author} {\bibfnamefont {S.}~\bibnamefont
  {Sachdev}}\ and\ \bibinfo {author} {\bibfnamefont {J.}~\bibnamefont {Ye}},\
  }\bibfield  {title} {\bibinfo {title} {Gapless spin-fluid ground state in a
  random quantum {Heisenberg} magnet},\ }\href@noop {} {\bibfield  {journal}
  {\bibinfo  {journal} {Physical Review Letters}\ }\textbf {\bibinfo {volume}
  {70}},\ \bibinfo {pages} {3339} (\bibinfo {year} {1993})}\BibitemShut
  {NoStop}%
\bibitem [{\citenamefont {Sachdev}(2010)}]{Sachdev2010}%
  \BibitemOpen
  \bibfield  {author} {\bibinfo {author} {\bibfnamefont {S.}~\bibnamefont
  {Sachdev}},\ }\bibfield  {title} {\bibinfo {title} {{Holographic Metals and
  the Fractionalized Fermi Liquid}},\ }\href
  {https://doi.org/10.1103/PhysRevLett.105.151602} {\bibfield  {journal}
  {\bibinfo  {journal} {Phys. Rev. Lett.}\ }\textbf {\bibinfo {volume} {105}},\
  \bibinfo {pages} {151602} (\bibinfo {year} {2010})}\BibitemShut {NoStop}%
\bibitem [{\citenamefont {Kitaev}(2015)}]{KitaevKITP}%
  \BibitemOpen
  \bibfield  {author} {\bibinfo {author} {\bibfnamefont {A.}~\bibnamefont
  {Kitaev}},\ }\bibfield  {title} {\bibinfo {title} {{A simple model of quantum
  holography}},\ }\href
  {http://online.kitp.ucsb.edu/online/entangled15/kitaev/} {\bibfield
  {journal} {\bibinfo  {journal} {Proc. KITP: Entanglement in
  Strongly-Correlated Quantum Matter 12}\ }\textbf {\bibinfo {volume} {26}}
  (\bibinfo {year} {2015})}\BibitemShut {NoStop}%
\bibitem [{\citenamefont {Maldacena}\ and\ \citenamefont
  {Stanford}(2016)}]{Maldacena2016}%
  \BibitemOpen
  \bibfield  {author} {\bibinfo {author} {\bibfnamefont {J.}~\bibnamefont
  {Maldacena}}\ and\ \bibinfo {author} {\bibfnamefont {D.}~\bibnamefont
  {Stanford}},\ }\bibfield  {title} {\bibinfo {title} {Remarks on the
  {Sachdev-Ye-Kitaev} model},\ }\href
  {https://doi.org/10.1103/PhysRevD.94.106002} {\bibfield  {journal} {\bibinfo
  {journal} {Phys. Rev. D}\ }\textbf {\bibinfo {volume} {94}},\ \bibinfo
  {pages} {106002} (\bibinfo {year} {2016})}\BibitemShut {NoStop}%
\bibitem [{\citenamefont {Banerjee}\ and\ \citenamefont
  {Altman}(2017{\natexlab{a}})}]{banerjeeSolvableModelDynamical2017c}%
  \BibitemOpen
  \bibfield  {author} {\bibinfo {author} {\bibfnamefont {S.}~\bibnamefont
  {Banerjee}}\ and\ \bibinfo {author} {\bibfnamefont {E.}~\bibnamefont
  {Altman}},\ }\bibfield  {title} {\bibinfo {title} {Solvable model for a
  dynamical quantum phase transition from fast to slow scrambling},\ }\href
  {https://doi.org/10.1103/PhysRevB.95.134302} {\bibfield  {journal} {\bibinfo
  {journal} {Physical Review B}\ }\textbf {\bibinfo {volume} {95}},\ \bibinfo
  {pages} {134302} (\bibinfo {year} {2017}{\natexlab{a}})}\BibitemShut
  {NoStop}%
\bibitem [{\citenamefont {Haldar}\ and\ \citenamefont
  {Shenoy}(2018)}]{Haldar2018c}%
  \BibitemOpen
  \bibfield  {author} {\bibinfo {author} {\bibfnamefont {A.}~\bibnamefont
  {Haldar}}\ and\ \bibinfo {author} {\bibfnamefont {V.~B.}\ \bibnamefont
  {Shenoy}},\ }\bibfield  {title} {\bibinfo {title} {{Strange half-metals and
  Mott insulators in Sachdev-Ye-Kitaev models}},\ }\href
  {https://doi.org/10.1103/PhysRevB.98.165135} {\bibfield  {journal} {\bibinfo
  {journal} {Physical Review B}\ }\textbf {\bibinfo {volume} {98}},\ \bibinfo
  {pages} {165135} (\bibinfo {year} {2018})}\BibitemShut {NoStop}%
\bibitem [{\citenamefont {Esterlis}\ and\ \citenamefont
  {Schmalian}(2019)}]{esterlisCooperPairingIncoherent2019}%
  \BibitemOpen
  \bibfield  {author} {\bibinfo {author} {\bibfnamefont {I.}~\bibnamefont
  {Esterlis}}\ and\ \bibinfo {author} {\bibfnamefont {J.}~\bibnamefont
  {Schmalian}},\ }\bibfield  {title} {\bibinfo {title} {Cooper pairing of
  incoherent electrons: {{An}} electron-phonon version of the
  {{Sachdev-Ye-Kitaev}} model},\ }\href
  {https://doi.org/10.1103/PhysRevB.100.115132} {\bibfield  {journal} {\bibinfo
   {journal} {Physical Review B}\ }\textbf {\bibinfo {volume} {100}},\ \bibinfo
  {pages} {115132} (\bibinfo {year} {2019})}\BibitemShut {NoStop}%
\bibitem [{\citenamefont
  {Wang}(2020)}]{wangSolvableStrongCouplingQuantumDot2020}%
  \BibitemOpen
  \bibfield  {author} {\bibinfo {author} {\bibfnamefont {Y.}~\bibnamefont
  {Wang}},\ }\bibfield  {title} {\bibinfo {title} {Solvable {{Strong-Coupling
  Quantum-Dot Model}} with a {{Non-Fermi-Liquid Pairing Transition}}},\ }\href
  {https://doi.org/10.1103/PhysRevLett.124.017002} {\bibfield  {journal}
  {\bibinfo  {journal} {Physical Review Letters}\ }\textbf {\bibinfo {volume}
  {124}},\ \bibinfo {pages} {017002} (\bibinfo {year} {2020})}\BibitemShut
  {NoStop}%
\bibitem [{\citenamefont {{Lantagne-Hurtubise}}\ \emph
  {et~al.}(2021)\citenamefont {{Lantagne-Hurtubise}}, \citenamefont {Pathak},
  \citenamefont {Sahoo},\ and\ \citenamefont {Franz}}]{lantagne2021}%
  \BibitemOpen
  \bibfield  {author} {\bibinfo {author} {\bibfnamefont {{\'E}.}~\bibnamefont
  {{Lantagne-Hurtubise}}}, \bibinfo {author} {\bibfnamefont {V.}~\bibnamefont
  {Pathak}}, \bibinfo {author} {\bibfnamefont {S.}~\bibnamefont {Sahoo}},\ and\
  \bibinfo {author} {\bibfnamefont {M.}~\bibnamefont {Franz}},\ }\bibfield
  {title} {\bibinfo {title} {Superconducting instabilities in a spinful
  {{Sachdev-Ye-Kitaev}} model},\ }\href
  {https://doi.org/10.1103/PhysRevB.104.L020509} {\bibfield  {journal}
  {\bibinfo  {journal} {Physical Review B}\ }\textbf {\bibinfo {volume}
  {104}},\ \bibinfo {pages} {L020509} (\bibinfo {year} {2021})}\BibitemShut
  {NoStop}%
\bibitem [{\citenamefont {Chowdhury}\ \emph {et~al.}(2022)\citenamefont
  {Chowdhury}, \citenamefont {Georges}, \citenamefont {Parcollet},\ and\
  \citenamefont {Sachdev}}]{Chowdhury2021b}%
  \BibitemOpen
  \bibfield  {author} {\bibinfo {author} {\bibfnamefont {D.}~\bibnamefont
  {Chowdhury}}, \bibinfo {author} {\bibfnamefont {A.}~\bibnamefont {Georges}},
  \bibinfo {author} {\bibfnamefont {O.}~\bibnamefont {Parcollet}},\ and\
  \bibinfo {author} {\bibfnamefont {S.}~\bibnamefont {Sachdev}},\ }\bibfield
  {title} {\bibinfo {title} {{Sachdev-Ye-Kitaev} models and beyond: Window into
  {non-Fermi} liquids},\ }\href {https://doi.org/10.1103/RevModPhys.94.035004}
  {\bibfield  {journal} {\bibinfo  {journal} {Rev. Mod. Phys.}\ }\textbf
  {\bibinfo {volume} {94}},\ \bibinfo {pages} {035004} (\bibinfo {year}
  {2022})}\BibitemShut {NoStop}%
\bibitem [{\citenamefont {Gu}\ \emph {et~al.}(2017)\citenamefont {Gu},
  \citenamefont {Qi},\ and\ \citenamefont {Stanford}}]{gu2017local}%
  \BibitemOpen
  \bibfield  {author} {\bibinfo {author} {\bibfnamefont {Y.}~\bibnamefont
  {Gu}}, \bibinfo {author} {\bibfnamefont {X.-L.}\ \bibnamefont {Qi}},\ and\
  \bibinfo {author} {\bibfnamefont {D.}~\bibnamefont {Stanford}},\ }\bibfield
  {title} {\bibinfo {title} {Local criticality, diffusion and chaos in
  generalized {Sachdev-Ye-Kitaev} models},\ }\href@noop {} {\bibfield
  {journal} {\bibinfo  {journal} {Journal of High Energy Physics}\ }\textbf
  {\bibinfo {volume} {2017}},\ \bibinfo {pages} {1} (\bibinfo {year}
  {2017})}\BibitemShut {NoStop}%
\bibitem [{\citenamefont {Haldar}\ \emph {et~al.}(2018)\citenamefont {Haldar},
  \citenamefont {Banerjee},\ and\ \citenamefont {Shenoy}}]{Haldar2018b}%
  \BibitemOpen
  \bibfield  {author} {\bibinfo {author} {\bibfnamefont {A.}~\bibnamefont
  {Haldar}}, \bibinfo {author} {\bibfnamefont {S.}~\bibnamefont {Banerjee}},\
  and\ \bibinfo {author} {\bibfnamefont {V.~B.}\ \bibnamefont {Shenoy}},\
  }\bibfield  {title} {\bibinfo {title} {{Higher-dimensional
  {Sachdev-Ye-Kitaev} {non-Fermi} liquids at {Lifshitz} transitions}},\ }\href
  {https://doi.org/10.1103/PhysRevB.97.241106} {\bibfield  {journal} {\bibinfo
  {journal} {Physical Review B}\ }\textbf {\bibinfo {volume} {97}},\ \bibinfo
  {pages} {1} (\bibinfo {year} {2018})}\BibitemShut {NoStop}%
\bibitem [{\citenamefont {Song}\ \emph {et~al.}(2017)\citenamefont {Song},
  \citenamefont {Jian},\ and\ \citenamefont {Balents}}]{Song2017a}%
  \BibitemOpen
  \bibfield  {author} {\bibinfo {author} {\bibfnamefont {X.~Y.}\ \bibnamefont
  {Song}}, \bibinfo {author} {\bibfnamefont {C.~M.}\ \bibnamefont {Jian}},\
  and\ \bibinfo {author} {\bibfnamefont {L.}~\bibnamefont {Balents}},\
  }\bibfield  {title} {\bibinfo {title} {{Strongly Correlated Metal Built from
  {Sachdev-Ye-Kitaev} Models}},\ }\href
  {https://doi.org/10.1103/PhysRevLett.119.216601} {\bibfield  {journal}
  {\bibinfo  {journal} {Physical Review Letters}\ }\textbf {\bibinfo {volume}
  {119}},\ \bibinfo {pages} {1} (\bibinfo {year} {2017})},\ \Eprint
  {https://arxiv.org/abs/1705.00117} {arXiv:1705.00117} \BibitemShut {NoStop}%
\bibitem [{\citenamefont {Zhang}(2017)}]{zhang2017}%
  \BibitemOpen
  \bibfield  {author} {\bibinfo {author} {\bibfnamefont {P.}~\bibnamefont
  {Zhang}},\ }\bibfield  {title} {\bibinfo {title} {Dispersive
  {Sachdev-Ye-Kitaev} model: Band structure and quantum chaos},\ }\href@noop {}
  {\bibfield  {journal} {\bibinfo  {journal} {Physical Review B}\ }\textbf
  {\bibinfo {volume} {96}},\ \bibinfo {pages} {205138} (\bibinfo {year}
  {2017})}\BibitemShut {NoStop}%
\bibitem [{\citenamefont {Chowdhury}\ \emph {et~al.}(2018)\citenamefont
  {Chowdhury}, \citenamefont {Werman}, \citenamefont {Berg},\ and\
  \citenamefont {Senthil}}]{Chowdhury2018}%
  \BibitemOpen
  \bibfield  {author} {\bibinfo {author} {\bibfnamefont {D.}~\bibnamefont
  {Chowdhury}}, \bibinfo {author} {\bibfnamefont {Y.}~\bibnamefont {Werman}},
  \bibinfo {author} {\bibfnamefont {E.}~\bibnamefont {Berg}},\ and\ \bibinfo
  {author} {\bibfnamefont {T.}~\bibnamefont {Senthil}},\ }\bibfield  {title}
  {\bibinfo {title} {Translationally invariant {Non-Fermi-Liquid} metals with
  critical {Fermi} surfaces: {{Solvable Models}}},\ }\href
  {https://doi.org/10.1103/PhysRevX.8.031024} {\bibfield  {journal} {\bibinfo
  {journal} {Phys. Rev. X}\ }\textbf {\bibinfo {volume} {8}},\ \bibinfo {pages}
  {031024} (\bibinfo {year} {2018})}\BibitemShut {NoStop}%
\bibitem [{\citenamefont {Jian}\ \emph {et~al.}(2017)\citenamefont {Jian},
  \citenamefont {Bi},\ and\ \citenamefont {Xu}}]{jian2017model}%
  \BibitemOpen
  \bibfield  {author} {\bibinfo {author} {\bibfnamefont {C.-M.}\ \bibnamefont
  {Jian}}, \bibinfo {author} {\bibfnamefont {Z.}~\bibnamefont {Bi}},\ and\
  \bibinfo {author} {\bibfnamefont {C.}~\bibnamefont {Xu}},\ }\bibfield
  {title} {\bibinfo {title} {Model for continuous thermal metal to insulator
  transition},\ }\href@noop {} {\bibfield  {journal} {\bibinfo  {journal}
  {Physical Review B}\ }\textbf {\bibinfo {volume} {96}},\ \bibinfo {pages}
  {115122} (\bibinfo {year} {2017})}\BibitemShut {NoStop}%
\bibitem [{\citenamefont {Aldape}\ \emph {et~al.}(2022)\citenamefont {Aldape},
  \citenamefont {Cookmeyer}, \citenamefont {Patel},\ and\ \citenamefont
  {Altman}}]{aldapeSolvableTheoryStrange2022}%
  \BibitemOpen
  \bibfield  {author} {\bibinfo {author} {\bibfnamefont {E.~E.}\ \bibnamefont
  {Aldape}}, \bibinfo {author} {\bibfnamefont {T.}~\bibnamefont {Cookmeyer}},
  \bibinfo {author} {\bibfnamefont {A.~A.}\ \bibnamefont {Patel}},\ and\
  \bibinfo {author} {\bibfnamefont {E.}~\bibnamefont {Altman}},\ }\bibfield
  {title} {\bibinfo {title} {Solvable {{Theory}} of a {{Strange Metal}} at the
  {{Breakdown}} of a {{Heavy Fermi Liquid}}},\ }\href@noop {} {\bibfield
  {journal} {\bibinfo  {journal} {arXiv:2012.00763}\ } (\bibinfo {year}
  {2022})},\ \Eprint {https://arxiv.org/abs/2012.00763} {arXiv:2012.00763}
  \BibitemShut {NoStop}%
\bibitem [{\citenamefont {Esterlis}\ \emph {et~al.}(2021)\citenamefont
  {Esterlis}, \citenamefont {Guo}, \citenamefont {Patel},\ and\ \citenamefont
  {Sachdev}}]{esterlis2021}%
  \BibitemOpen
  \bibfield  {author} {\bibinfo {author} {\bibfnamefont {I.}~\bibnamefont
  {Esterlis}}, \bibinfo {author} {\bibfnamefont {H.}~\bibnamefont {Guo}},
  \bibinfo {author} {\bibfnamefont {A.~A.}\ \bibnamefont {Patel}},\ and\
  \bibinfo {author} {\bibfnamefont {S.}~\bibnamefont {Sachdev}},\ }\bibfield
  {title} {\bibinfo {title} {Large- {N} theory of critical {Fermi} surfaces},\
  }\href {https://doi.org/10.1103/PhysRevB.103.235129} {\bibfield  {journal}
  {\bibinfo  {journal} {Physical Review B}\ }\textbf {\bibinfo {volume}
  {103}},\ \bibinfo {pages} {235129} (\bibinfo {year} {2021})}\BibitemShut
  {NoStop}%
\bibitem [{\citenamefont {Georges}\ \emph {et~al.}(1996)\citenamefont
  {Georges}, \citenamefont {Kotliar}, \citenamefont {Krauth},\ and\
  \citenamefont {Rozenberg}}]{Georges1996}%
  \BibitemOpen
  \bibfield  {author} {\bibinfo {author} {\bibfnamefont {A.}~\bibnamefont
  {Georges}}, \bibinfo {author} {\bibfnamefont {G.}~\bibnamefont {Kotliar}},
  \bibinfo {author} {\bibfnamefont {W.}~\bibnamefont {Krauth}},\ and\ \bibinfo
  {author} {\bibfnamefont {M.~J.}\ \bibnamefont {Rozenberg}},\ }\bibfield
  {title} {\bibinfo {title} {Dynamical mean-field theory of strongly correlated
  fermion systems and the limit of infinite dimensions},\ }\href
  {https://doi.org/10.1103/RevModPhys.68.13} {\bibfield  {journal} {\bibinfo
  {journal} {Rev. Mod. Phys.}\ }\textbf {\bibinfo {volume} {68}},\ \bibinfo
  {pages} {13} (\bibinfo {year} {1996})}\BibitemShut {NoStop}%
\bibitem [{\citenamefont {Held}(2007)}]{Held2007}%
  \BibitemOpen
  \bibfield  {author} {\bibinfo {author} {\bibfnamefont {K.}~\bibnamefont
  {Held}},\ }\bibfield  {title} {\bibinfo {title} {Electronic structure
  calculations using dynamical mean field theory},\ }\href
  {https://doi.org/10.1080/00018730701619647} {\bibfield  {journal} {\bibinfo
  {journal} {Advances in Physics}\ }\textbf {\bibinfo {volume} {56}},\ \bibinfo
  {pages} {829} (\bibinfo {year} {2007})},\ \Eprint
  {https://arxiv.org/abs/https://doi.org/10.1080/00018730701619647}
  {https://doi.org/10.1080/00018730701619647} \BibitemShut {NoStop}%
\bibitem [{\citenamefont {Haldar}\ \emph {et~al.}(2021)\citenamefont {Haldar},
  \citenamefont {Tavakol},\ and\ \citenamefont {Scaffidi}}]{HaldarPRR2020}%
  \BibitemOpen
  \bibfield  {author} {\bibinfo {author} {\bibfnamefont {A.}~\bibnamefont
  {Haldar}}, \bibinfo {author} {\bibfnamefont {O.}~\bibnamefont {Tavakol}},\
  and\ \bibinfo {author} {\bibfnamefont {T.}~\bibnamefont {Scaffidi}},\
  }\bibfield  {title} {\bibinfo {title} {Variational wave functions for
  {Sachdev-Ye-Kitaev} models},\ }\href
  {https://doi.org/10.1103/PhysRevResearch.3.023020} {\bibfield  {journal}
  {\bibinfo  {journal} {Phys. Rev. Research}\ }\textbf {\bibinfo {volume}
  {3}},\ \bibinfo {pages} {023020} (\bibinfo {year} {2021})}\BibitemShut
  {NoStop}%
\bibitem [{\citenamefont {Kotliar}\ \emph {et~al.}(2001)\citenamefont
  {Kotliar}, \citenamefont {Savrasov}, \citenamefont {P\'alsson},\ and\
  \citenamefont {Biroli}}]{Kotliar2001}%
  \BibitemOpen
  \bibfield  {author} {\bibinfo {author} {\bibfnamefont {G.}~\bibnamefont
  {Kotliar}}, \bibinfo {author} {\bibfnamefont {S.~Y.}\ \bibnamefont
  {Savrasov}}, \bibinfo {author} {\bibfnamefont {G.}~\bibnamefont
  {P\'alsson}},\ and\ \bibinfo {author} {\bibfnamefont {G.}~\bibnamefont
  {Biroli}},\ }\bibfield  {title} {\bibinfo {title} {Cellular dynamical mean
  field approach to strongly correlated systems},\ }\href
  {https://doi.org/10.1103/PhysRevLett.87.186401} {\bibfield  {journal}
  {\bibinfo  {journal} {Phys. Rev. Lett.}\ }\textbf {\bibinfo {volume} {87}},\
  \bibinfo {pages} {186401} (\bibinfo {year} {2001})}\BibitemShut {NoStop}%
\bibitem [{\citenamefont {Capone}\ and\ \citenamefont
  {Kotliar}(2006)}]{Capone2006}%
  \BibitemOpen
  \bibfield  {author} {\bibinfo {author} {\bibfnamefont {M.}~\bibnamefont
  {Capone}}\ and\ \bibinfo {author} {\bibfnamefont {G.}~\bibnamefont
  {Kotliar}},\ }\bibfield  {title} {\bibinfo {title} {Competition between
  $d$-wave superconductivity and antiferromagnetism in the two-dimensional
  hubbard model},\ }\href {https://doi.org/10.1103/PhysRevB.74.054513}
  {\bibfield  {journal} {\bibinfo  {journal} {Phys. Rev. B}\ }\textbf {\bibinfo
  {volume} {74}},\ \bibinfo {pages} {054513} (\bibinfo {year}
  {2006})}\BibitemShut {NoStop}%
\bibitem [{\citenamefont {Patel}\ and\ \citenamefont
  {Sachdev}(2019)}]{patelTheoryPlanckianMetal2019}%
  \BibitemOpen
  \bibfield  {author} {\bibinfo {author} {\bibfnamefont {A.~A.}\ \bibnamefont
  {Patel}}\ and\ \bibinfo {author} {\bibfnamefont {S.}~\bibnamefont
  {Sachdev}},\ }\bibfield  {title} {\bibinfo {title} {Theory of a {{Planckian}}
  metal},\ }\href {https://doi.org/10.1103/PhysRevLett.123.066601} {\bibfield
  {journal} {\bibinfo  {journal} {Physical Review Letters}\ }\textbf {\bibinfo
  {volume} {123}},\ \bibinfo {pages} {066601} (\bibinfo {year} {2019})},\
  \Eprint {https://arxiv.org/abs/1906.03265} {arXiv:1906.03265} \BibitemShut
  {NoStop}%
\bibitem [{\citenamefont {Jose}\ \emph {et~al.}(2022)\citenamefont {Jose},
  \citenamefont {Seo},\ and\ \citenamefont
  {Uchoa}}]{joseNonFermiLiquidBehavior2022}%
  \BibitemOpen
  \bibfield  {author} {\bibinfo {author} {\bibfnamefont {G.}~\bibnamefont
  {Jose}}, \bibinfo {author} {\bibfnamefont {K.}~\bibnamefont {Seo}},\ and\
  \bibinfo {author} {\bibfnamefont {B.}~\bibnamefont {Uchoa}},\ }\bibfield
  {title} {\bibinfo {title} {Non-fermi liquid behavior in the sachdev-ye-kitaev
  model for a one-dimensional incoherent semimetal},\ }\href
  {https://doi.org/10.1103/PhysRevResearch.4.013145} {\bibfield  {journal}
  {\bibinfo  {journal} {Phys. Rev. Research}\ }\textbf {\bibinfo {volume}
  {4}},\ \bibinfo {pages} {013145} (\bibinfo {year} {2022})}\BibitemShut
  {NoStop}%
\bibitem [{\citenamefont {Maharaj}\ \emph {et~al.}(2017)\citenamefont
  {Maharaj}, \citenamefont {Rosenberg}, \citenamefont {Hristov}, \citenamefont
  {Berg}, \citenamefont {Fernandes}, \citenamefont {Fisher},\ and\
  \citenamefont {Kivelson}}]{maharajTransverseFieldsTune2017}%
  \BibitemOpen
  \bibfield  {author} {\bibinfo {author} {\bibfnamefont {A.~V.}\ \bibnamefont
  {Maharaj}}, \bibinfo {author} {\bibfnamefont {E.~W.}\ \bibnamefont
  {Rosenberg}}, \bibinfo {author} {\bibfnamefont {A.~T.}\ \bibnamefont
  {Hristov}}, \bibinfo {author} {\bibfnamefont {E.}~\bibnamefont {Berg}},
  \bibinfo {author} {\bibfnamefont {R.~M.}\ \bibnamefont {Fernandes}}, \bibinfo
  {author} {\bibfnamefont {I.~R.}\ \bibnamefont {Fisher}},\ and\ \bibinfo
  {author} {\bibfnamefont {S.~A.}\ \bibnamefont {Kivelson}},\ }\bibfield
  {title} {\bibinfo {title} {Transverse fields to tune an {{Ising-nematic}}
  quantum phase transition},\ }\href {https://doi.org/10.1073/pnas.1712533114}
  {\bibfield  {journal} {\bibinfo  {journal} {Proceedings of the National
  Academy of Sciences}\ }\textbf {\bibinfo {volume} {114}},\ \bibinfo {pages}
  {13430} (\bibinfo {year} {2017})}\BibitemShut {NoStop}%
\bibitem [{\citenamefont {Gastiasoro}\ \emph {et~al.}(2014)\citenamefont
  {Gastiasoro}, \citenamefont {Paul}, \citenamefont {Wang}, \citenamefont
  {Hirschfeld},\ and\ \citenamefont {Andersen}}]{Andersen_defect2014}%
  \BibitemOpen
  \bibfield  {author} {\bibinfo {author} {\bibfnamefont {M.~N.}\ \bibnamefont
  {Gastiasoro}}, \bibinfo {author} {\bibfnamefont {I.}~\bibnamefont {Paul}},
  \bibinfo {author} {\bibfnamefont {Y.}~\bibnamefont {Wang}}, \bibinfo {author}
  {\bibfnamefont {P.~J.}\ \bibnamefont {Hirschfeld}},\ and\ \bibinfo {author}
  {\bibfnamefont {B.~M.}\ \bibnamefont {Andersen}},\ }\bibfield  {title}
  {\bibinfo {title} {Emergent defect states as a source of resistivity
  anisotropy in the nematic phase of iron pnictides},\ }\href
  {https://doi.org/10.1103/PhysRevLett.113.127001} {\bibfield  {journal}
  {\bibinfo  {journal} {Phys. Rev. Lett.}\ }\textbf {\bibinfo {volume} {113}},\
  \bibinfo {pages} {127001} (\bibinfo {year} {2014})}\BibitemShut {NoStop}%
\bibitem [{\citenamefont {Marciani}\ and\ \citenamefont
  {Benfatto}(2022)}]{Benfatto2022}%
  \BibitemOpen
  \bibfield  {author} {\bibinfo {author} {\bibfnamefont {M.}~\bibnamefont
  {Marciani}}\ and\ \bibinfo {author} {\bibfnamefont {L.}~\bibnamefont
  {Benfatto}},\ }\bibfield  {title} {\bibinfo {title} {Resistivity anisotropy
  from the multiorbital {Boltzmann} equation in nematic {FeSe}},\ }\href
  {https://doi.org/10.1103/PhysRevB.106.045102} {\bibfield  {journal} {\bibinfo
   {journal} {Phys. Rev. B}\ }\textbf {\bibinfo {volume} {106}},\ \bibinfo
  {pages} {045102} (\bibinfo {year} {2022})}\BibitemShut {NoStop}%
\bibitem [{\citenamefont {Banerjee}\ and\ \citenamefont
  {Altman}(2017{\natexlab{b}})}]{Banerjee2017b}%
  \BibitemOpen
  \bibfield  {author} {\bibinfo {author} {\bibfnamefont {S.}~\bibnamefont
  {Banerjee}}\ and\ \bibinfo {author} {\bibfnamefont {E.}~\bibnamefont
  {Altman}},\ }\bibfield  {title} {\bibinfo {title} {{Solvable model for a
  dynamical quantum phase transition from fast to slow scrambling}},\
  }\bibfield  {journal} {\bibinfo  {journal} {Physical Review B}\ }\textbf
  {\bibinfo {volume} {95}},\ \href {https://doi.org/10.1103/PhysRevB.95.134302}
  {10.1103/PhysRevB.95.134302} (\bibinfo {year} {2017}{\natexlab{b}}),\ \Eprint
  {https://arxiv.org/abs/1610.04619} {arXiv:1610.04619} \BibitemShut {NoStop}%
\bibitem [{\citenamefont {Sahoo}\ \emph {et~al.}(2020)\citenamefont {Sahoo},
  \citenamefont {{Lantagne-Hurtubise}}, \citenamefont {Plugge},\ and\
  \citenamefont {Franz}}]{sahoo2020}%
  \BibitemOpen
  \bibfield  {author} {\bibinfo {author} {\bibfnamefont {S.}~\bibnamefont
  {Sahoo}}, \bibinfo {author} {\bibfnamefont {{\'E}.}~\bibnamefont
  {{Lantagne-Hurtubise}}}, \bibinfo {author} {\bibfnamefont {S.}~\bibnamefont
  {Plugge}},\ and\ \bibinfo {author} {\bibfnamefont {M.}~\bibnamefont
  {Franz}},\ }\bibfield  {title} {\bibinfo {title} {Traversable wormhole and
  {{Hawking-Page}} transition in coupled complex {{SYK}} models},\ }\href
  {https://doi.org/10.1103/PhysRevResearch.2.043049} {\bibfield  {journal}
  {\bibinfo  {journal} {Physical Review Research}\ }\textbf {\bibinfo {volume}
  {2}},\ \bibinfo {pages} {043049} (\bibinfo {year} {2020})}\BibitemShut
  {NoStop}%
\bibitem [{\citenamefont {{Garc{\'i}a-Garc{\'i}a}}\ \emph
  {et~al.}(2021)\citenamefont {{Garc{\'i}a-Garc{\'i}a}}, \citenamefont
  {Zheng},\ and\ \citenamefont {Ziogas}}]{garcia2021}%
  \BibitemOpen
  \bibfield  {author} {\bibinfo {author} {\bibfnamefont {A.~M.}\ \bibnamefont
  {{Garc{\'i}a-Garc{\'i}a}}}, \bibinfo {author} {\bibfnamefont {J.~P.}\
  \bibnamefont {Zheng}},\ and\ \bibinfo {author} {\bibfnamefont
  {V.}~\bibnamefont {Ziogas}},\ }\bibfield  {title} {\bibinfo {title} {Phase
  diagram of a two-site coupled complex {{SYK}} model},\ }\href
  {https://doi.org/10.1103/PhysRevD.103.106023} {\bibfield  {journal} {\bibinfo
   {journal} {Physical Review D}\ }\textbf {\bibinfo {volume} {103}},\ \bibinfo
  {pages} {106023} (\bibinfo {year} {2021})}\BibitemShut {NoStop}%
\end{thebibliography}%

\onecolumngrid

\appendix

\renewcommand{\thesection}{{SI~\arabic{section}}}
\renewcommand{\theequation}{\thesection.\arabic{equation}}
\renewcommand{\theequation}{SI~\arabic{equation}}
\renewcommand{\thefigure}{SI~\arabic{figure}}

\section{Lattice Model Free Energy Derivation}
In this section, we derive the effective self-consistent equations for the disorder averaged two-orbital lattice SYK model in the $N \mapsto \infty$ limit.  
We start with the Hamiltonian (Eq. 1, 2, and 3 of the main text), given by 
\begin{align}
\label{Ham}
    H_{\rm kin} &=  \sum_{\bk,s,i} \varepsilon^\pdg_s(\bk) c^{\dg}_{\bk,s,i} c^{\pdg}_{\bk,s,i} \\
    H^{\rm intra}_{\text{SYK}}  &=  \sum_{\br,s,(ijkl)} 
   J^{(s)}_{ijkl}(\br) 
  c_{\br,s,i}^{\dagger} c_{\br,s,j}^{\dagger} c^\pdg_{\br,s,k} 
  c^\pdg_{\br,s,l} \\
H^{\rm inter}_{\text{SYK}} &=
  \sum_{\br,(ijkl)} V_{ijkl}^\pdg(\br) 
  c_{\br,+, i}^{ \dagger}  c^{ \dagger}_{\br,+,j} 
  c^\pdg_{\br,-,k}  c^\pdg_{\br,-,l} + {\mathrm {h.c.}}
\end{align}

In order to construct the corresponding action, we utilize the replica trick for disorder averaging the partition function ($Z$)
\begin{equation}
    \beta \overline{F} = - \overline{ \ln (Z)} = - \lim_{M  \mapsto 0} \partial_M \overline{Z^M} = - \lim_{M  \mapsto 0} \frac{\overline{Z^M}-1}{M},
\end{equation}
where there are $M$ replicas (copies) of the system and $N$ fermionic SYK modes for each copy. In the large-$N$ limit, $\overline{Z^M}$ can be separated if we assume off-diagonal correlations between different copies fall off as $\frac{1}{N}$ \cite{Chowdhury2021b}
\begin{equation}
\label{ansatz}
  \overline{Z^M} = \overline{Z}^M + \mathcal{O}\left(\frac{1}{N}\right).
\end{equation}
This separation can be implemented via a  `replica diagonal ansatz' \cite{Banerjee2017b} which expresses  $\overline{Z}$ as
\begin{equation}
\begin{gathered}
\label{partition}
\overline{Z} = \int  \prod_s \mathcal{D}(\bar{c}, c)
\exp \left[ -  \int d\tau_1  \sum_{i, \br, \br^\prime} \bar{c}_{s,i, \br} \left(\partial_{\tau_1} - \mu + t_s(\br - \br^\prime) \right) c_{s,i,\br^\prime} \right. \\
- \frac{N J^{2}}{4} \int d\tau_1d\tau_2  \left(\frac{1}{N^2} \sum_{i,j,\br} \bar{c}_{s,i,\br}(\tau_1)c_{s,i,\br}(\tau_2) \bar{c}_{s,j,\br}(\tau_2)  c_{s,j,\br}(\tau_1) \right)^{2} \\
- \left. \frac{N V^2  }{4} \int d\tau_1d\tau_2 \left( \frac{1}{N^2}\sum_{i,j,\br} \bar{c}_{s, i,\br}(\tau_1) c_{s, i,\br} (\tau_2)  \bar{c}_{-s,j,\br}(\tau_2)c_{-s,j,\br} (\tau_1)\right) ^2\right] ,
\end{gathered}
\end{equation}
where $t_s(\br - \br^\prime)$ denotes the hopping amplitude between $\br, \br^\prime$ for each orbital (see Fig. 1 in the main text), with the Fourier transform corresponding to $\varepsilon_s(\bk)$ in Eq. \ref{Ham}. The other symbols are defined near Eq. 1, 2, and 3 of the main text. The disorder average of $J,V$ contained a $\delta_{\br,\br^\prime}$ term, which simplified the spatial sums to arrive at Eq. \ref{partition}.  Next, we introduce the bilocal fields $G_s$ via a $\delta$-function identity,  equivalent to introducing the Lagrange multipliers specifying $\Sigma_s$. 
Introducing the  Lagrange multipliers ($\Sigma_s$) corresponds to adding the action 
\begin{equation}
S_{\Sigma}=-N\int_{0}^{\beta} \sum_{\br} d\tau_1d\tau_2 \Sigma_s\left(\tau_1,\tau_2, \br \right)\left[ G_s\left(\tau_2,  \tau_1,\br \right)- \frac{1}{N}\sum_{i=1}^{N} \overline{c}_{s,i}(\tau_1,\br ) c_{s,i}\left(\tau_2,\br\right)\right]
\end{equation}
so that interaction terms, such as the $J^2$ term in Eq.\ \ref{partition},  can be expressed as
\begin{equation}
\begin{gathered}
\exp[-\frac{N J^{2}}{4} \sum_\br \int d\tau_1 d\tau_2  \left(\frac{1}{N} \sum_{i=1}^{N} \bar{c}_{s,i,\br}(\tau_1) c_{s,i,\br}  (\tau_2 )\right)^{2} \left(\frac{1}{N} \sum_{i=1}^{N}\bar{c}_{s,i,\br}(\tau_2) c_{s,i,\br}  (\tau_1 )\right)^{2}] = \\
\int \mathcal{D} G_s \mathcal{D} \Sigma_s \exp \left[ -N \sum_\br \int d\tau_1d\tau_2  \bigg( \frac{J^{2}}{4} G_s  (\tau_1,\tau_2,\br )^{2}G_s  (\tau_2,\tau_1, \br )^2 \right. \\
+\left. \Sigma_s  (\tau_1,\tau_2, \br )   G_s  (\tau_2,\tau_1,\br )-\Sigma_s  (\tau_1,\tau_2, \br ) \frac{1}{N} \sum_{i=1}^{N} \bar{c}_{s,i,\br}(\tau_1) c_{s,i,\br}  (\tau_2 ) \bigg) \right].
\end{gathered}
\end{equation}
A similar transformation can also be done for the $V^2$ term in \eq{partition} as well.  This transformation expresses $\overline{Z}$ in terms of an effective action $S$ which is quadratic in the fermion fields $\overline{c}, c$ and parameterized by $G_s$ and $\Sigma_s$. Appealing to translation invariance of the disorder averaged system, $\Sigma_s$ and $G_s$ are same for each lattice site ($\br$) and therefore momentum ($\bk$) independent.The action $S$ can then be expressed in the band basis,   as
\begin{equation}
\begin{gathered}
    S =  -  \int d\tau_{1,2} \sum_{s,\bk,i} \bar{c}_{s,\bk,i} (\tau_1) \bigl( \partial_\tau \delta(\tau_1 - \tau_2) - \mu + \varepsilon_s(\bk) + \Sigma_s (\tau_1, \tau_2) \bigr) c_{s,\bk,i}(\tau_2) -  N \mathcal{V} \int d\tau_{1,2} \Sigma_s  (\tau_1,\tau_2 ) G_s  (\tau_2,\tau_1 ) \\
    -   N \mathcal{V} \int d\tau_{1,2} \biggl( \frac{J^{2} }{4} G_s^{2}(\tau_{1}, \tau_{2}  ) G_s^{2}(\tau_{2}, \tau_{1}) + \frac{V^2}{4} G^2_{-s}(\tau_{1}, \tau_{2})G_s^2(\tau_{2}, \tau_{1}) \biggr)
\end{gathered}
\end{equation}
with a resulting dispersion 
\begin{equation}
\varepsilon_s(\bk) = -2 t_s\left(\cos k_{x} a +\cos k_{y} a \right)-2 \delta t_s \left(\cos k_{x} a-\cos k_{y} a\right) ,
\end{equation}
where $\delta t_s = (-1)^s \delta t$ parameterizes the symmetric anisotropy and $\sum_\br = \mathcal{V}$. 
We trace over the fermionic degrees of freedom to obtain the free energy functional per fermionic mode ($N$) per volume ($\mathcal{V}$) 
\begin{equation}
\begin{gathered}
\Omega =- \frac{1}{\mathcal{V}} \sum_s  \operatorname{Tr} \ln [(\partial_{\tau_{1}}- \mu +\varepsilon_s(\bk)  ) \delta(\tau_{1}-\tau_{2}  )+\Sigma_s(\tau_{1}, \tau_{2}  )  ] 
-  \\
\int \mathrm{d} \tau_{1,2}\sum_s \left[ \frac{J^{2} }{4} G_s^{2}(\tau_{1}, \tau_{2}  ) G_s^{2}(\tau_{2}, \tau_{1} )  
 + \Sigma_s(\tau_{1}, \tau_{2} ) G_s(\tau_{2}, \tau_{1}  )  \right.
\left.+ \frac{V^2}{4} G^2_{-s}(\tau_{1}, \tau_{2})G_s^2(\tau_{2}, \tau_{1}) \right].
\end{gathered}
\end{equation}
Demanding imaginary time translation invariance, i.e., $G_s\left(\tau_{1}, \tau_{2}\right)=G_s\left(\tau_{1}-\tau_{2} = \tau \right) $ and same for $\Sigma_s$, we can simplify $\Omega$ to  
\begin{equation}
\label{eq:freeenergy}
\begin{gathered}
\Omega =\sum_{s=\pm}\left[- \frac{1}{\beta} \sum_{i \omega_{n}} 
\int \mathrm{d} \varepsilon g_s(\varepsilon) \ln \left[\mathfrak{i} \omega_{n}  +\mu-\varepsilon-\Sigma_s\left(\mathfrak{i} \omega_{n}\right)\right] +\int_{0}^{\beta} \mathrm{d} \tau \Sigma_s(\tau) G_s(\beta-\tau) - \frac{J^{2}}{4} \int_{0}^{\beta} \mathrm{d} \tau 
G_s^{2}(\beta-\tau) G_s^{2}(\tau)\right]  \\
-  \frac{V^2}{2} \int_{0}^{\beta} d \tau G_{+}^{2}(\beta-\tau) G_{-}^{2}(\tau)
\end{gathered}
\end{equation}
where $g_s(\varepsilon) = \int \frac{d^2 {\bk}}{(2\pi)^2} \delta(\varepsilon-\varepsilon_s(\bk))$ 
is the lattice density of states for orbital-$s$,   $\omega_n\!=\! (2 n \!+\! 1)\pi/\beta$ 
represents the fermionic Matsubara frequencies, and $\beta=T^{-1}$ with $T$ denoting the temperature. The above expression appears in Eq.\ 4 of the main text.\par
Next we extremize $\Omega$ by demanding $\delta_G \Omega = \delta_\Sigma \Omega = 0$ to find a saddle-point described by the Dyson equations
\begin{equation}
\begin{gathered}
\Sigma_s(\tau )= -J^{2} G_s^{2}(\tau) G_s(- \tau)
-V^2 G_{-s}^{2}(\tau) G_{s}( - \tau), 
\\
G_s(\tau) = - \frac{1}{\mathcal{V}} \int_{BZ} d\bk \  \left[ \partial_\tau - \mu + \varepsilon_s(\bk) + \Sigma_s\right]^{-1} .
\end{gathered}
\end{equation}
The equivalent Matsubara local Green's function is given by (Eq.\ 6 in the main text)
\begin{equation}
\label{eq:green_mat}
    G_s(i\omega_n)= \int \mathrm{d} \varepsilon \ g_s(\varepsilon) \left[i\omega_n+\mu-\varepsilon-\Sigma_s(i\omega_n)\right]^{-1} . 
\end{equation}
\section{Spectral Function Calculation}

In this section, we detail the calculation for numerical analytic continuation of $G_s(i \omega_n)$ in order to determine the spectral function $A_s(\omega)$.  We obtain the spectral functions through a complementary self-consistent approach in order to determine the unique spectral properties in each phase and to compute transport quantities. We follow the approach as established in \cite{Banerjee2017b,Haldar2018b}. 
We determine $\Sigma^R_s(\omega)$ through the Fourier transformed Dyson equation
\begin{equation}
\begin{gathered}
\Sigma_s\left(i \omega_{n}\right)=\int_{0}^{\beta} d \tau e^{i \omega_{n} \tau} \Sigma_s(\tau)=-\frac{J^{2}}{\beta^{2}} \sum_{n_{1}, n_{2}} G_s\left(i \omega_{n_{1}}\right) G_s\left(i \omega_{n_{2}}\right) G_s\left(i \omega_{n_{1}}-i \omega_{n_{2}}+i \omega_{n}\right)  \\
+ \frac{V^{2}}{\beta^{2}} \sum_{n_{1}, n_{2}} G_s\left(i \omega_{n_{1}}\right) G\left(i \omega_{n_{2}}\right) G_{-s}\left(i \omega_{n_{1}}-i \omega_{n_{2}}+i \omega_{n}\right)
\end{gathered}
\end{equation}
Now we express the spectral representation $G_s(i \omega_n) = \int d \omega \frac{A_s(\omega)}{(i \omega_n - \omega)} $, where the first term (per flavour $s$) for example is written as
\begin{equation}
\begin{gathered}
   - \sum_{n_1, n_2}\frac{J^2}{\beta^2} \int_{-\infty}^{\infty}  \frac{\left( \prod \limits ^3_{\alpha = 1}  d \omega_\alpha A_s(\omega_\alpha) \right)}{(i \omega_{n_1} - \omega_1)(i \omega_{n_2} - \omega_2)(i \omega_{n_1} + i \omega_{n_2} + i \omega_n - \omega_3)} .
\end{gathered}
\end{equation}
The Matsubara sums in the expression are given as the residue of the simple poles 
\begin{equation}
\begin{gathered}
  \frac{1}{\beta}\sum_{i \omega_{n_2} }  \left(\lim_{z \mapsto \omega_1} \frac{n_{\rm F}(z)}{(i \omega_{n_2} -\omega_2)(z + i \omega_{n_2}  - \omega_3)}+ \lim_{z \mapsto \omega_3  - i \omega_n - i \omega_{n_2}}\frac{n_{\rm F}(z)}{(z  - \omega_1)(i \omega_{n_2} -\omega_2)} \right) \\
  = \frac{1}{\beta} \sum_{i \omega_{n_2} \mapsto z_2} \frac{n_{\rm F}(\omega_1) - n_{\rm F}(\omega_3   + i \omega_n -i \omega_{n_2})}{(z_2 -\omega_2)(z_2 -(\omega_3 - \omega_1 - i \omega_n))}\\
  = \left(n_{\rm F}(\omega_1) - n_{\rm F}(\omega_3)\right)\left(\lim_{z_2 \mapsto \omega_2} \frac{n_{\rm F}(z)}{(\omega_2 -\omega_3 + \omega_1 + i \omega_n)}+ \lim_{z_2 \mapsto \omega_3 - \omega_{1} - i \omega_n }\frac{n_{\rm F}(z)}{(\omega_3 - \omega_1 - i \omega_n -\omega_2)} \right) \\
  = \left(n_{\rm F}(\omega_1) - n_{\rm F}(\omega_3)\right)  \frac{\left( n_{\rm F}(\omega_2) + n_B(\omega_3 - \omega_1) \right) }{ \left(i \omega_n + \omega_1 +\omega_2 - \omega_3  \right)},
\end{gathered}
\end{equation}
where sums of combinations of Matsubara frequencies transmute fermionic and bosonic distribution functions. Combining the distribution functions and analytically continuing $i \omega_n \to \omega + i \eta $ gives 
\begin{equation}
\begin{gathered}
        -J^2  \int_{-\infty}^{\infty}  \left( \prod \limits ^3_{\alpha = 1}  d \omega_\alpha A_s(\omega_\alpha) \right) \left(n_{\rm F}(\omega_1) - n_{\rm F}(\omega_3)\right) \frac{n_{\rm F}(\omega_2) + n_B(\omega_3 - \omega_1) }{\omega_1 +\omega_2 - \omega_3 - i \omega_n} \\
        = J^2  \int_{-\infty}^{\infty}  \left( \prod \limits ^3_{\alpha = 1}  d \omega_\alpha A_s(\omega_\alpha) \right)  \frac{\left( n_{\rm F}(-\omega_1)n_{\rm F}(-\omega_2)n_{\rm F}(\omega_3) + n_{\rm F}(\omega_1)n_{\rm F}(\omega_2)n_{\rm F}(-\omega_3)  \right)}{(\omega + \omega_1  + \omega_2 - \omega_3 + i \eta) } . 
\end{gathered}
\end{equation}
A similar expression can be derived for the interorbital $V$ term. Combining both $J$ and $V$ terms and using the identity $\frac{1}{\omega^{+}}=-i \int_{0}^{\infty} d t \exp (i \omega t)$, gives
\begin{equation}
\begin{aligned}
\Sigma_{s}\left(\omega^{+}\right)=&-i \int_{0}^{\infty} d t e^{-i \omega t}\left[J^{2}\left\{n^2_{1 s}(t) n_{2 s}(t)+n^2_{3 s}(t) n_{4 s}(t)\right\}\right.\\
&\left.+ V^{2}\left\{n^2_{1 -s}(t) n_{2, s}(t)+n^{2}_{3 -s}(t) n_{4, s}(t)\right\}\right]
\end{aligned}
\end{equation}
where the time-dependent occupations are integrated from $[-\infty, +\infty]$ as 
\begin{equation}
\begin{aligned}
n_{1 s}(t)&=\int d \omega A_s\left(\omega\right) n_{\rm F}\left(-\omega\right) e^{i \omega t} \\
n_{2 s}(t)&=\int d \omega A_s\left(\omega\right) n_{\rm F}\left(\omega\right) e^{-i \omega t}  \\
n_{3 s}(t)&=\int d \omega A_s\left(\omega\right) n_{ \rm F}\left(\omega\right) e^{i \omega t} \\
n_{4 s}(t)&=\int d \omega A_s\left(\omega\right) n_{\rm F}\left(-\omega\right) e^{-i \omega t} .
\end{aligned}
\end{equation}
The corresponding Dyson equation via analytic continuation is
\begin{equation}
G^R_{s}( \omega)= \int \mathrm{d}  \varepsilon \ g(\varepsilon) \left[\omega +\mu- \varepsilon- \Sigma^R_{s}\left(\omega \right)\right]^{-1}
\end{equation}
which we use to numerically compute 
\begin{equation}
    A_s( \omega) = \frac{-1}{\pi} \operatorname{Im}\left(  G^R_s(\omega ) \right).
\end{equation}
\section{Additional Thermodynamic Investigations}
\label{si_thermo}
\begin{figure}[t]
    \centering
\includegraphics[width=1\textwidth]{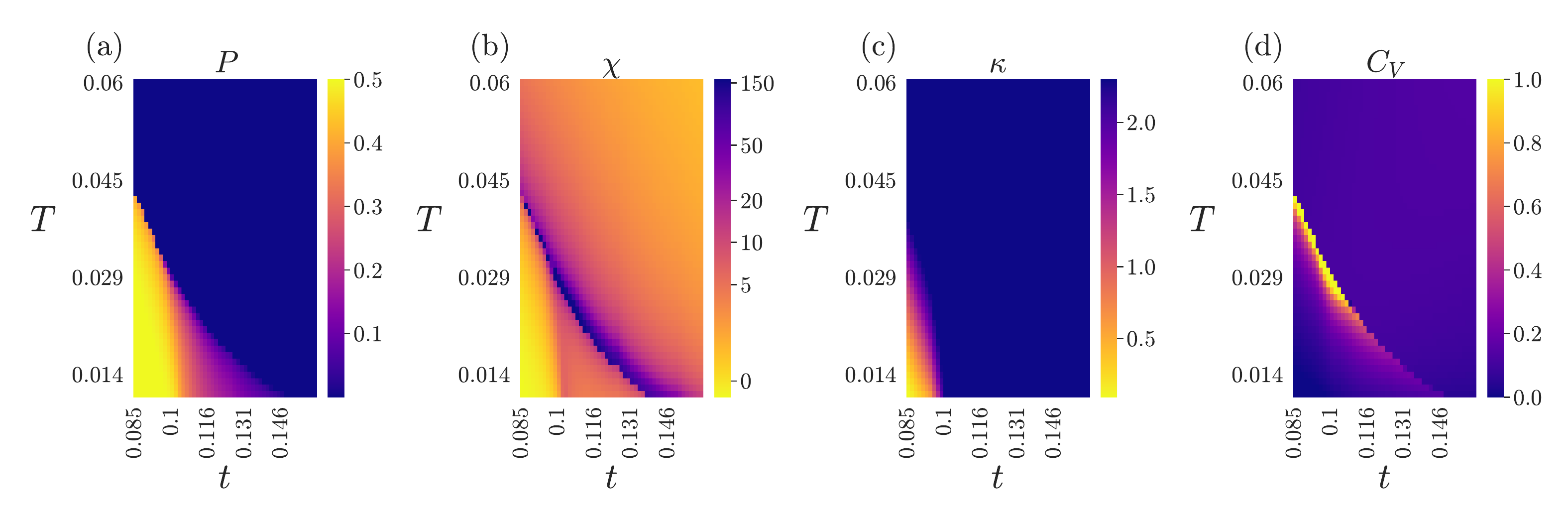}
    \caption{Thermodynamic signatures of the phase transition. $T$ is temperature and $t$ is hopping. (a) gives $P$, the polarization. (b) gives the susceptibility $\chi$, in $\text{log}$ scale,$(c)$ gives $S$, the entropy, and $(d)$ gives $C_V$, the specific heat. }
    \label{fig:supp_phase}
\end{figure}

In this section, we show the additional signatures of the nematic phase transition in order to characterize the nature of the transition. 
As described in the phase diagram section, the polarization $P =  \langle n_+ 
\rangle - \langle n_{-} \rangle$, (shown in Fig. \ref{fig:supp_phase}(a)), tracks both first and second order transitions. These transitions are also visible in the (dis)continuity in entropy $S = - \partial\Omega/\partial T$ at the  (first) second order transition. This feature corresponds to an amplified height of the peaks in specific heat $C_v =T \partial S / \partial T$ (shown in Fig.\ \ref{fig:supp_phase}(b)). \par
\begin{figure}[t]
    \centering
\includegraphics[width=1\textwidth]{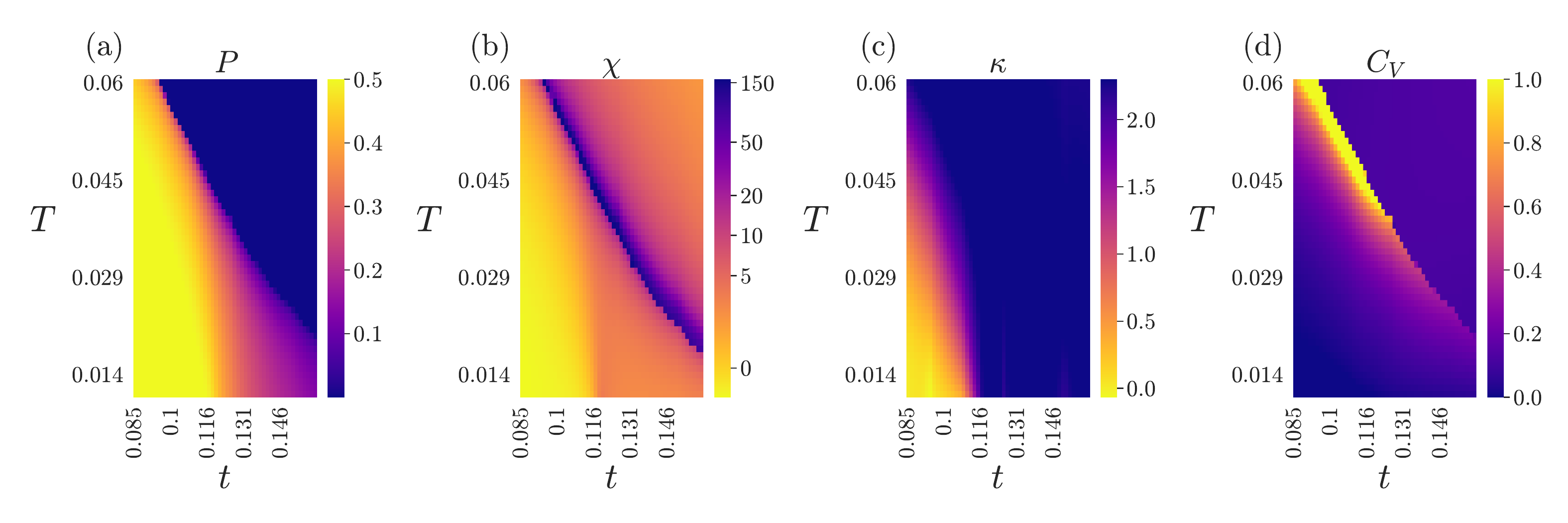}
    \caption{Thermodynamic signatures of the phase transition with $V = 1.1$. $T$ is temperature and $t$ is hopping. (a) gives $P$, the polarization. (b) gives the susceptibility $\chi$, in $\text{log}$ scale,$(c)$ gives $S$, the entropy, and $(d)$ gives $C_V$, the specific heat. The transition line moves towards higher temperatures and higher hoppings. }
    \label{fig:supp_2_phase}
\end{figure}
\begin{figure}[h!]
    \centering
\includegraphics[width=1\textwidth]{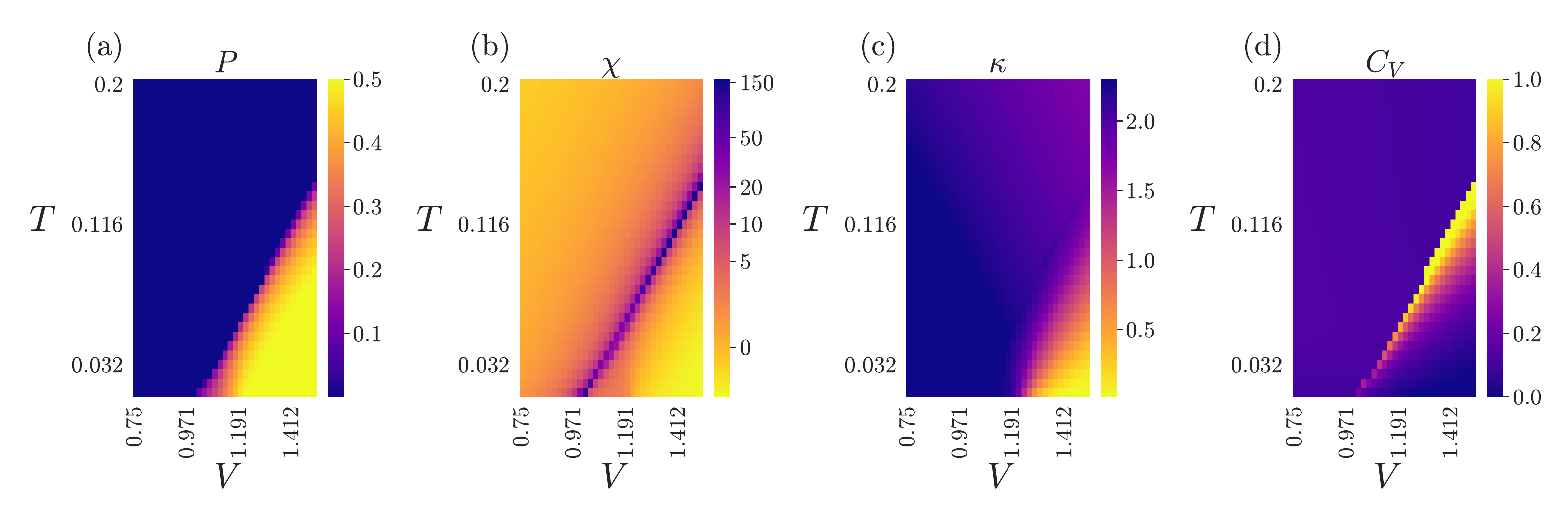}
    \caption{Thermodynamic signatures of the phase transition with tuning inter-orbital interaction strength ($V$). $T$ is temperature and $t$ is hopping. (a) gives $P$, the polarization. (b) gives the susceptibility $\chi$, in $\text{log}$ scale,$(c)$ gives $S$, the entropy, and $(d)$ gives $C_V$, the specific heat. Here $t = 0.13$  and $\delta t / t \approx 0.05 $}
    \label{fig:supp_V}
\end{figure}

In addition, we consider the susceptibility $\chi = \partial\Omega/\partial h |_{h = 0} $, by considering a small transverse field (or strain) that couples to each orbital ($s = \pm $).  The modified bare Green's function describes this coupling as $G_{s}(i \omega_n) = \int d \varepsilon  \ g(\varepsilon)  (i \omega_n +  s h + \mu - \varepsilon)^{-1}$. $\chi$ shows a strong peak (shown in $\text{log}$ in Fig.\ \ref{fig:supp_phase}(d)) along  the transition. $\chi$ also distinguishes between the NI and NM phases by dropping to a lower value in the NI phase.  We also demonstrate the effect of tuning the inter-orbital interaction strength $V$. (shown in Fig.\ \ref{fig:supp_V}). There is a critical $V \sim 1 $ in which this polarization-symmetry breaking is suppressed below this value. Increasing $V > 1$  leads to an eventual change in the location of the phase boundary, where the transition occurs at higher temperatures as $V$ increases. This means that the tricrtical point shifts, where the first-order transition continues for larger $t$. This is also demonstrated by a representative plot for $V = 1.1$ shown in Fig.\ \ref{fig:supp_2_phase}.
\section{ Momentum Resolved Spectral Functions and Mass Enhancement}

We present the momentum resolved spectral functions $A_s(\bk,\omega)$ for the $+$ and $-$ orbitals (along a path in the BZ connecting the high-symmetry momenta $\Gamma$, $X$, $M$) in \Fig{fig:spectral} (a) and (b), respectively. The combined spectral function $A(\bk,\omega)=\sum_{s=\pm}A_s(\bk,\omega)$ should be measurable using photoemission spectroscopic techniques such as angle-resolved photoemission spectroscopy (ARPES). The  Fermi-surface (FS) for the respective orbitals are also shown as insets of \Fig{fig:spectral}(a) and (b). We determine these Fermi-surfaces by solving for the set of momenta $\{k_F\}$ that show peaks for $A(\bk_F,\omega=0)$.  

We now mention some key features of the FS in the nematic-metal phase (\Fig{fig:spectral}) that are distinct from the FS obtained in the isotropic-metallic phase.  First, the density imbalance between the orbitals generated in the nematic phase causes a change in the FS geometry. This change in geometry splits a FS degeneracy at $(\pi/2, \pi/2)$ momentum present in the isotropic phase and shifts the new surfaces to different momenta points. Therefore, when we move along the $\Gamma-M$ line (see \Fig{fig:spectral} insets), we cross the FS twice (once for each orbital flavor) in the nematic metal as opposed to once for the isotropic phase. Second, for the $\Gamma-X$ and $X-M$ lines, while the number of FS crossings remains the same for the nematic and isotropic phases, the locations for the crossings shift significantly in the nematic-metal phase. Hence, in addition to our d.c. transport predictions presented in the main text, the above features of the momentum resolved spectral function should prove helpful in detecting the onset of a nematic-metal phase in experiments.

In addition to the momentum-resolved analysis, we consider the electronic mass-enhancement within the FL and nFL phases, as discussed in the main text. These quantities serve as indications of the degree of correlations within the system.

As $\pdv{\Sigma_s(\bk, \omega)}{\bk} = 0$ in our model, we define the mass enhancement( $m^*(T) / m$) and quasiparticle residue $Z^{-1}(T)$ as 
\begin{equation}
    m^*(T) / m  = Z^{-1}(T) = \left(1- \lim_{\omega \to 0} \pdv{\Sigma(\omega)}{\omega}\right) 
\end{equation}
We reiterate that we drop the orbital label since these observables are the same for both orbitals. 
We are interested in the $Z(T \!\to \! 0)$ behaviour, which is the limit in which these observables are strictly defined. This corresponds to the true quasiparticle
residue in a FL ground state. In comparison, this quantity is ill-defined, or divergent, for nFL states. The divergence of $Z^{-1}$ can define a nFL state. \par
\begin{figure}[h!]
    \centering
\includegraphics[width=0.8\textwidth]{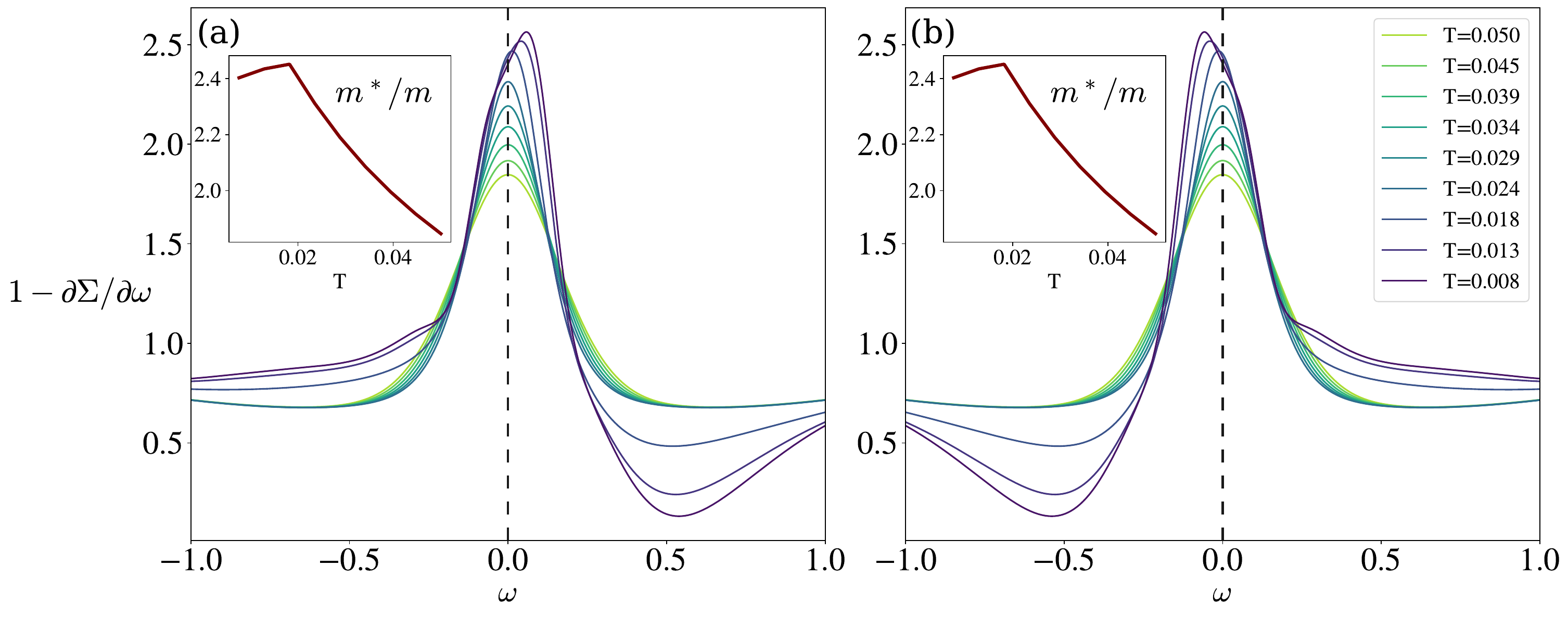}
    \caption{Plot of $1 - \pdv{\Sigma}{\omega}$, where the limit $\omega \mapsto 0$ corresponds to the quasiparticle residue $Z$ and in this model, the mass enhancement $m/m^*$. Here $t = 0.11$ and $\delta t / t = 0.05$.  $Z^{-1}$ increases as $T$ is lowered until the phase transition occurs and the system begins to behave as a Fermi liquid. Then $Z^{-1}$ ceases to increase with decreasing $T$ and saturates at a finite value.  }
    \label{fig:massenhancement}
\end{figure}
\begin{figure}[h!]
    \centering
\includegraphics[width=0.8\textwidth]{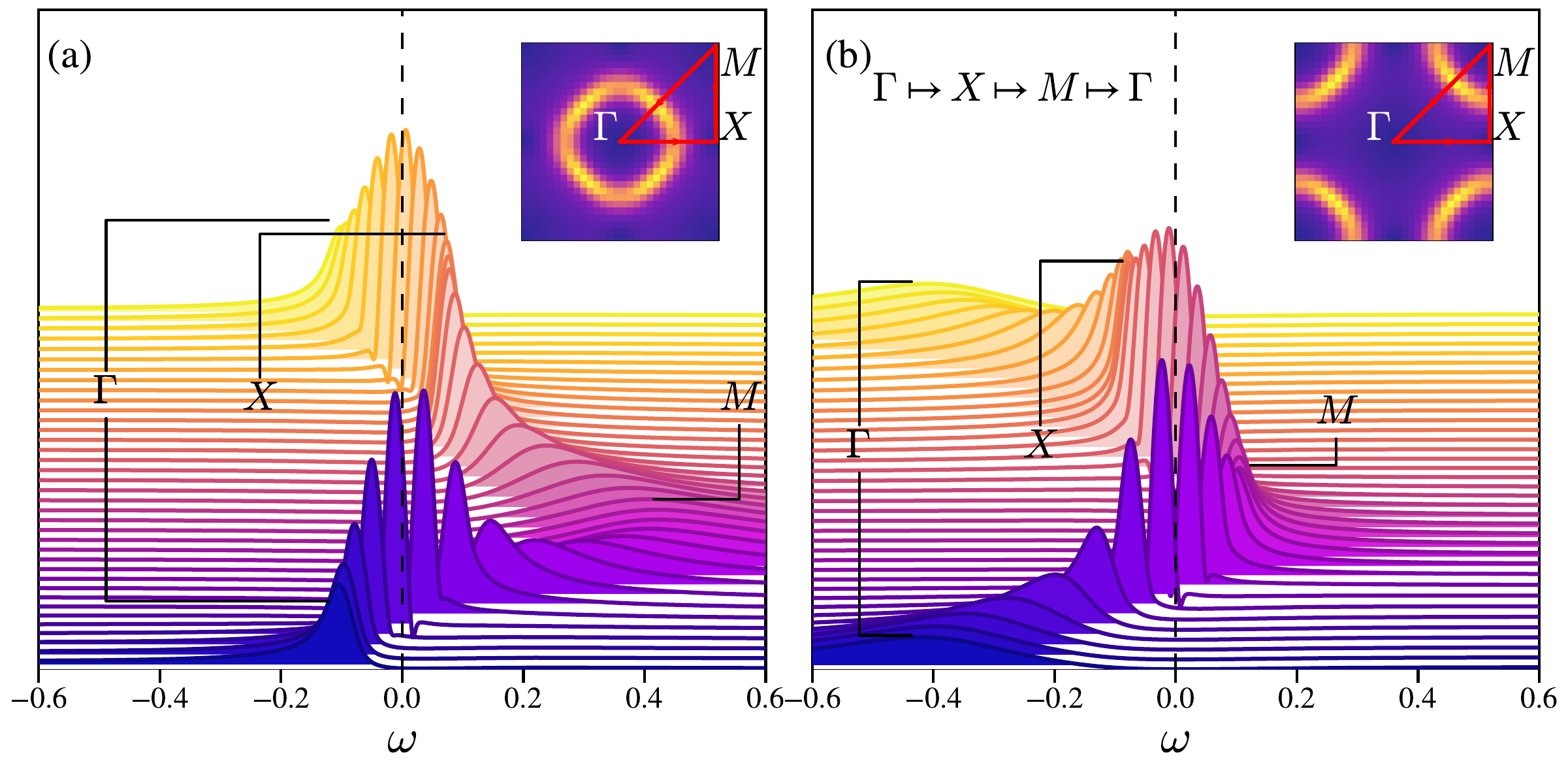}
    \caption{Plot of the momentum ($\bk$) resolved spectral function $A_s(\bk,\omega)$ for the $+$(a) and $-$(b) orbital following a high-symmetry path through the Brillouin Zone $(\Gamma \to X \to M \to \Gamma)$ demonstrating the Fermi surface (inset) in the nematic-metal (NM) phase.  Here the hopping is $t = 0.11$, anisotropy is $\delta t / t = 0.05$ and  the temperature is $T = 0.016$ respectively.  (The anisotropy of the Fermi surface shape in the nematic phase is not obviously evident in this plot due to the small scale of $\delta t / t$).}
    \label{fig:spectral}
\end{figure}

As illustrated in Fig. \ref{fig:massenhancement}, the mass enhancement from our numerical results in the nFL regime indicate an increasing $m^*/m$ for each orbital upon lowering temperature, indicative of a divergent $Z^{-1}$ in the $T \to 0$ limit. $\left(1- \pdv{\Sigma(\omega)}{\omega}\right) $ shows signatures of the nematic transition for finite $\omega$ through a spectral asymmetry at low frequency. 
For $\omega = 0$ however, the two orbitals display identical behaviour. The divergent $Z^{-1}$ in the $T \to 0$ limit is consistent with the nFL definition. The increasing $Z^{-1}(T)$ ceases below the nematic transition, leading to a large but finite mass enhancement, and a corresponding reduced $Z$. 

\section{Transverse Strain}
In this section, we examine the effect that additional forms of strain have on the model. In order to determine the tunability of the tricritical phase transition, we consider on-site $B_{2g}$ strain, which couples 
to the off-diagonal orbital elements as
$H_{\gamma} =  \gamma \sum_{s} c^\dagger_{s} c_{-s} $. 

\begin{figure}[t]
    \centering
\includegraphics[width=0.8\textwidth]{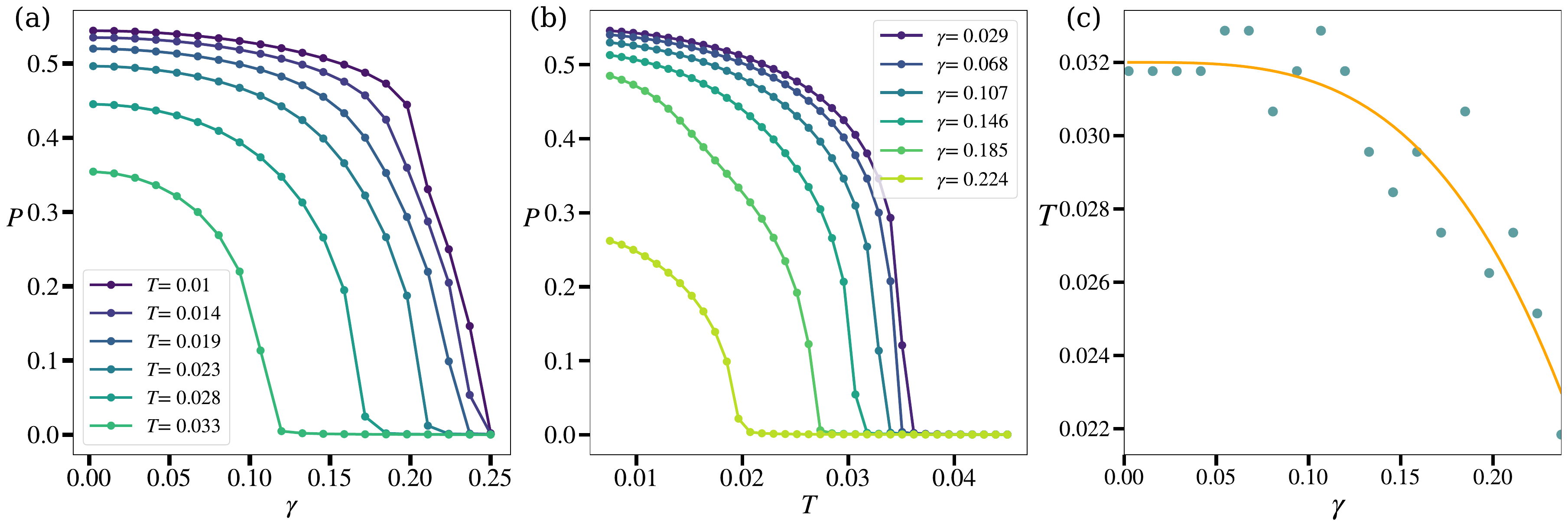}
    \caption{Suppression of transition at  $\delta t/t=0.05$. For a representative hopping point in the continuous transition ($t = 0.11$), (a) shows the vanishing polarization at each $T$ value with increasing strain $\gamma$. Increasing $\gamma$ corresponds to reducing $T_c$ for the transition. (b) shows the suppressed polarization symmetry breaking for specific $\gamma$ values. (c) demonstrates the decreasing trend in the tricritical $T_c$ with increasing transverse strain $\gamma$~; the fluctuations are numerical artifacts.}
    \label{fig:strain}
\end{figure}

In corresponding momentum space, it gives an inverse Matsubara Green's function
\begin{equation}
    G^{-1}(\bk, i \omega_n) = \begin{pmatrix}
i \omega_n - \varepsilon_{s}(\bk) & \gamma \\
\gamma & i \omega_n - \varepsilon_{-s}(\bk)
\end{pmatrix}
+ 
\begin{pmatrix}
\Sigma_{s}( i \omega_n) & 0 \\
0 & \Sigma_{-s}( i \omega_n)
\end{pmatrix}
.
\end{equation}
By inverting the $G^{-1}$ matrix and using the local Green's function  $G(i\omega_n) = \frac{1}{\mathcal{V}}\int d \bk \ G( \bk, i \omega _n)$, we obtain the following form for the local Green's function 
\begin{equation}
    G_s(i \omega_n) = \int_{\text{BZ}} d \bk \frac{\left(i\omega_n - \varepsilon_{-s}(\bk) - \Sigma_{-s}(i \omega_n) \right)}{\left(i\omega_n - \varepsilon_{s}(\bk) - \Sigma_{s}(i \omega_n)\right)\left(i\omega_n - \varepsilon_{-s}(\bk) - \Sigma_{-s}(i \omega_n)\right)-\gamma^2}.
    \label{transverse}
\end{equation}
Eq. \ref{transverse} requires additional computational time compared to \eq{eq:green_mat} (Eq.~6 main text) as its evaluation requires performing an explicit integral over the full Brillouin zone, rather than an effective integral over $\varepsilon$ weighted by the DOS ($g(\varepsilon$)). We follow the same procedure detailed in \cite{Haldar2018b, Haldar2018c}. This momentum integral is done in a discretized $40 \times 40$ point grid and each parameter space point takes from $ \approx 100$ to $\approx 6000$ iterations to converge depending on the proximity to the transition. 
The self consistent solutions show a strong suppression of the nematic ordering to lower temperatures (Fig. \ref{fig:strain} (a) and (b)).  This pushes the tricritical point to lower values (Fig. \ref{fig:strain} (c)). More detailed numerical studies could examine the possibility of this value approaching numerical zero at a putative quantum critical point.  

\section{Competing Phases}
In order to explore possible alternative ordered phases, we consider an enlarged unit cell that can include inter-site correlations.  The  isotropic hopping Hamiltonian then becomes

\begin{equation}
 H_0(\vec{k}) = -2 t    \begin{bmatrix}
0 & \cos(k_x a) & \cos(k_y a) & 0 \\
\cos(k_x a) & 0 & 0 & \cos(k_y a) \\
\cos(k_y a) & 0 & 0 & \cos(k_x a) \\
0 &  \cos(k_y a) & \cos(k_x a) & 0 
\end{bmatrix}  
\end{equation}
This lattice has been used to explore potential $d-$wave ordering and anti-ferromagnetic order in the Hubbard model \cite{Kotliar2001,Capone2006}. 
The corresponding multi-site Dyson equations are expressed in terms of matrix elements as 
\begin{equation}
    \label{Full dyson equations inter}
    \begin{gathered}
        (G^{0})^{-1}_{\alpha\beta}(\tau, \vec{k}) = \delta^{\alpha\beta}\left(\partial_{\tau}-\mu_0\right)+H_0^{\alpha\beta}(\vec{k})\,,\\
        G_{\alpha\beta}(\tau_1-\tau_2)=-\frac{1}{\mathcal{V}}\sum_{\vec{k}}\left(\left(\delta(\tau_2-\tau_1)(G^0)^{-1}(\tau_1, \vec{k}) +\Sigma(\tau_1-\tau_2)\right)^{-1}_{\alpha\beta}\right)\,,\\
        \Sigma_{\alpha\alpha}(\tau) = -\frac{1}{4}(J)^2G^2_{\alpha\alpha}(\tau)G_{\alpha\alpha}(-\tau) \,.
    \end{gathered}
\end{equation}
From these runs with finite anisotropy $\delta t$ and finite hopping $t$, we determine two competing solutions with comparable free-energies. 
One such solution to these Dyson equations gives translationally invariant results, where the polarization $P$ is the same magnitude and sign for each of the four sites. This corresponds to the ferronematic order.  
By considering initial conditions with an alternating solution, the system converges to a `checkerboard' solution where diagonal sites have matching $P$, with an opposite sign to neighboring sites. This checkerboard corresponds to an antiferronematic order.  
\par
\begin{figure}[t]
    \centering
\includegraphics[width=0.8\textwidth]{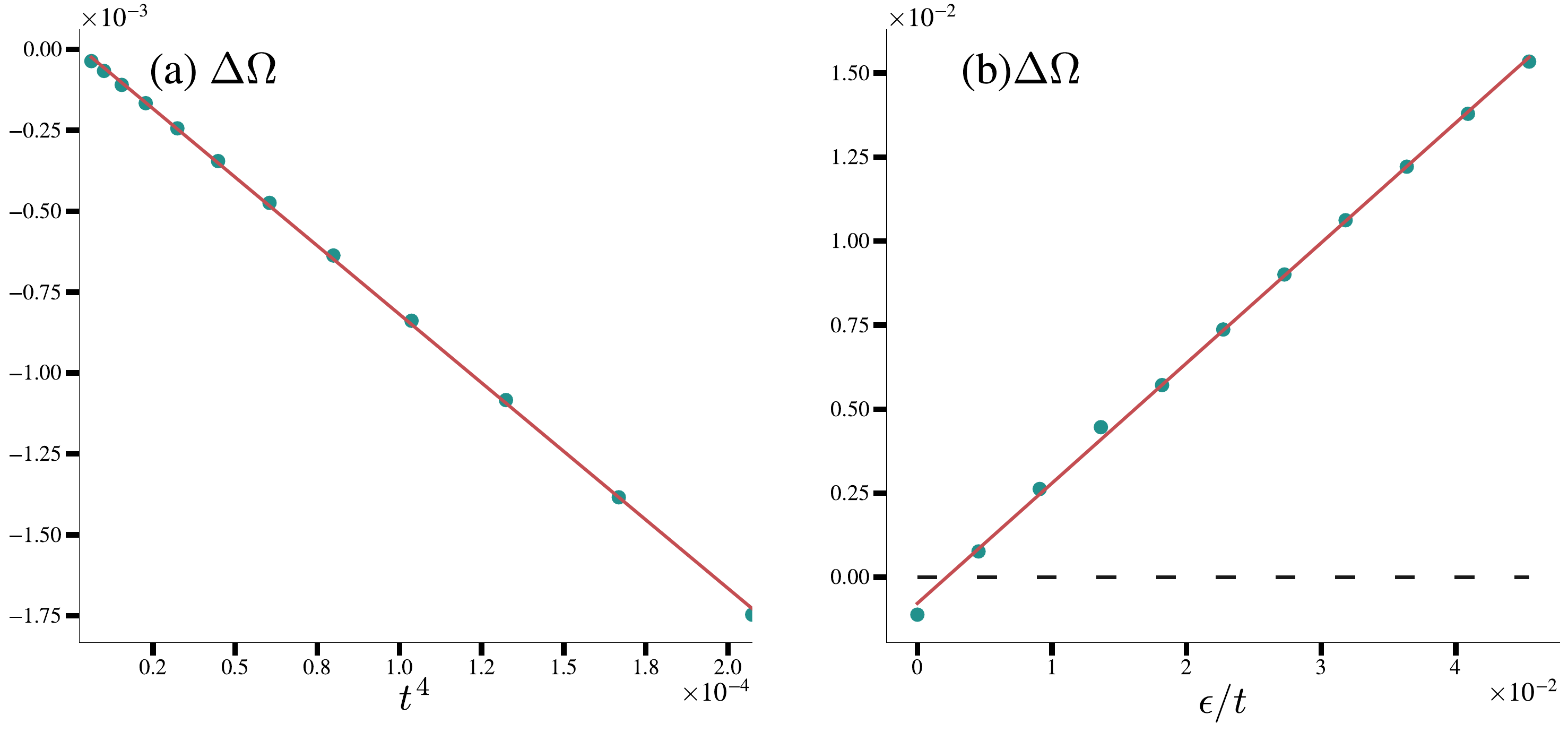}
    \caption{A plot of the difference in free-energy densities between the ferronematic state and the antiferronematic state ($\Delta \Omega =  \Omega_a - \Omega_f$). a) Demonstrates a $t^4 /V^3 \in [0.05,0.12]$ scaling with $\Delta \Omega$ (red line) with a coefficient $ \sim 10$. The parameters are   $\delta t / t = 0.05, V = J = 1, T = 0.02, \epsilon = 0 $.  b) Increasing $B1g$ strain ($\epsilon$) gives a linear response that increases $\Delta_\Omega$ (red fitted line).  Here a representative point in parameter space was chosen with $t = 0.11, \delta t / t = 0.05, V = J = 1, T = 0.02$.  }
    \label{fig:groundstate}
\end{figure}

 
The difference in free-energy density between the ferronematic and antiferronematic state is $\Delta \Omega =  \Omega_a - \Omega_f$.
Throughout the phase diagram, the two states are nearly degenerate with one another ($ \Delta \Omega  \propto t^4/V^3$), 
with the antiferronematic state having a {\it very} slightly lower free energy. 
The hopping dependency on $\Delta \Omega$ (\ref{fig:groundstate}(a)) demonstrates that this competition occurs 
at a higher order than the naive expectation that exchange interactions that might drive ordering at order $t^2 / V$. 
This might suggest an `electron pair hopping' mechanism for exchange interactions between SYK dots which we are currently investigating further.
Applying even a tiny $B1g$ strain field ($\epsilon /t \sim  10^{-3}$) lifts this near-degeneracy
and induces a net uniform polarization ($P$), which favours the translationally invariant ferronematic state as the eventual ground-state (Fig. \ref{fig:groundstate}(b)).

\section{Nature of the Phase Transition: A Phenomenological Perspective}
The inter-orbital interaction term
\begin{equation}
    -\frac{V^2}{2} \int d\tau\ G_+(\beta-\tau)^2G_-(\tau)^2 
\end{equation}
in the free energy functional (Eq.~4 of main text) facilitates the spontaneous symmetry breaking in our model and appears as a direct consequence of keeping only pair-hopping terms between the orbitals (see Eq.~3, main text).
The inter-orbital term above and others in Eq.~4 are symmetric under the particle-hole (PH) transformation $G_s(\tau)\to G_s(\beta-\tau)$, which effectively amounts to a reflection about $\tau=\beta/2$ in the imaginary time. While at high temperatures, our model prefers a PH symmetric solution $G_s(\tau)=G_s(\beta-\tau)$  (see \Fig{fig:LG_panel}(a)), at lower temperatures, the $\int d\tau\ G_+(\beta-\tau)^2 G_-(\tau)^2$ term favors a PH-asymmetric solution $G_s(\tau)\neq G_s(\beta-\tau)$ since the latter minimizes the free energy.  As a result, we encounter a spontaneous breaking of the PH symmetry when $T\to0$, which occurs via a first-order or second-order phase transition depending on the hopping parameters. \par

\def\drho{\delta\rho}
\begin{figure}[t]
    \centering
\includegraphics[width=0.8\textwidth]{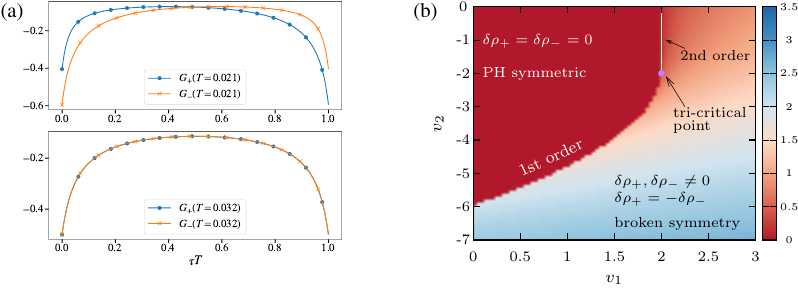}
    \caption{(a) Imaginary-time Green's functions $G_{+}(\tau)$ (circles), $G_{-}(\tau)$ (crosses) for the two orbitals ($+$, $-$) plotted as function of $\tau T$, where $T$ is the temperature. The top figure shows the solutions obtained by minimizing the free energy breaks particle-hole (PH) symmetry at lower temperatures. In comparison, the bottom figure shows the PH symmetric solutions obtained for higher temperatures. (b) Phase diagram obtained by minimizing a Landau-Ginzburg energy functional used to describe the effective theory of coupled SYK orbitals. The symbols $\drho_+$, $\drho_-$ represent the density deviations away from half-filling for each orbital, and $v_1$, $v_2$ are the couplings between the two orbitals (see text for details).}
    \label{fig:LG_panel}
\end{figure}

We can gain a qualitative understanding of how the thermal symmetry-breaking transition gets tuned between first and second order as 
follows --
The PH-symmetric solution fixes the density of fermions ($\rho_{s=\pm}$) in each orbital to $0.5$;  this can be shown using the identities 
\begin{align}
    \rho_s=&G_s(\tau=0^-),\nonumber\\  G_s(\tau=0^-)-&G_s(\tau=0^+)=1\nonumber,\\ G_s(\tau)=&-G_s(\beta+\tau)\nonumber
\end{align} for fermionic Green's functions and the symmetry condition $G_s(\tau)=G_s(\beta-\tau)$. When PH symmetry breaks, although the total density remains the same, the fermion density in each orbital deviates from $0.5$. Therefore, we can quantify the degree of PH-symmetry breaking associated with each orbital using the deviations $\drho_s=\rho_s-0.5$ and use them to explain the nature of the transition.

To do so, we adopt a phenomenological semiclassical Landau-Ginzburg (LG) approach, which is a good starting point for exploring thermal phase transitions,
and minimize the following energy functional 
\begin{equation}\label{eq:LG1}
\Omega(\drho_+,\drho_-) = \left[\sum_{s=\pm} a_2 (\drho_s)^2 + a_4 (\drho_s)^4 + (\drho_s)^6\right] + v_1\drho_+\drho_- + v_2(\drho_+\drho_-)^2,
\end{equation}
where $v_1$, $v_2$ represent couplings between the orbitals and $a_2$, $a_4$ are parameters describing the effective theory of the individual orbitals. The $(\drho_s)^6$ term ensures $\Omega>0$ for large values of $\drho_s$ and is required for the stability of the LG theory.  Under the PH-symmetry operation, $\drho_s$ transforms as $\drho_s \to -\drho_s$, implying that the LG functional $\Omega(\drho_+,\drho_-)$ is PH symmetric like the free energy of our model given in Eq.~4 of the main text. Interestingly, the LG functional above is reminiscent of an effective theory of coupled Ising-like models. The latter is known to undergo first or second order transitions to a broken symmetry phase. Therefore, the functional $\Omega$ for our model also demonstrates similar features for a broad range of parameters. E.g., for $a_2=a_4=1$, we find a phase diagram in the $v_1$--$v_2$ plane (\Fig{fig:LG_panel}(b)) showing the existence of first and second order transitions that meet at a critical point, thus capturing the same qualitative behavior seen in our exact numerical calculations (Fig.~2(a) and (c) of the main text). Looking more closely, we note that the solutions in both the symmetry broken and unbroken phases are of the form $\drho_+ =-\drho_- \equiv \varphi$. Rewriting the LG functional (\eq{eq:LG1}) in terms of $\varphi$ we find an effective functional 
\begin{equation}
    \Omega(\varphi) = (2 a_2 - v_1) \varphi^2 + (2 a_4 - v_2) \varphi^4 + 2\varphi^6
\end{equation}
where $v_1$ renormalizes $a_2 \to a_2 - v_1 /2 $ and $v_2$ renormalizes $a_4 \to a_4-v_2 /2$. The functional above clearly resembles a typical ($\varphi^2$, $\varphi^4$, $\varphi^6$) theory where the coefficient of the quartic ($\varphi^4$) term can lead to a first-order transition when $(2a_4-v_2)<0$. The above arguments demonstrate that such transitions, governed by symmetry considerations, are a general feature of our multiorbital SYK model.

Similar first and  second order transitions have also been reported in previous works ~\cite{Haldar2018c, sahoo2020, garcia2021, lantagne2021} that studied coupled SYK models defined on isolated quantum dots \emph{without} a lattice. E.g., ref.~\cite{Haldar2018c} which explored a model of two coupled SYK dots, also explains these transitions using a more sophisticated version of the LG functional argument presented above by incorporating quantum fluctuations. However, we emphasize that these previous studies \emph{did not} connect such a transition to a  nematic order \emph{on a lattice} and focused only on the physics of isolated quantum dots. 
\end{document}